\documentclass[11pt]{elsarticle}

\usepackage[margin=2.5cm]{geometry}
\usepackage{amsmath}
\usepackage{amssymb}
\usepackage{amsfonts}
\usepackage{upgreek}
\usepackage{here} 
\usepackage[figurename=Fig.,labelsep=period]{caption}
\usepackage[subrefformat=parens]{subcaption}
\usepackage{color}
\usepackage{comment}
\usepackage{algorithm}
\usepackage{algpseudocode}
\usepackage{multirow}
\usepackage{hyperref}
\hypersetup{
    colorlinks=true,       
    linkcolor=blue,        
    citecolor=blue,        
    filecolor=blue,        
    urlcolor=blue          
}

\begin{document}
	\begin{frontmatter}
		\title{Latent Space-based Stochastic Model Updating}
		
\author[inst1]{Sangwon Lee\corref{cor1}}
\author[inst1]{Taro Yaoyama}
\author[inst2]{Masaru Kitahara}
\author[inst1]{Tatsuya Itoi}

\affiliation[inst1]{organization={The University of Tokyo, Department of Architecture},
	addressline={7-3-1, Hongo}, city={Bunkyo-ku}, 
	state={Tokyo}, country={Japan}}
\affiliation[inst2]{organization={The University of Tokyo, Department of Civil Engineering},
	addressline={7-3-1, Hongo}, city={Bunkyo-ku},
	state={Tokyo}, country={Japan}}

\cortext[cor1]{Corresponding author. \\ \hspace*{1.5em} \textit{Email address}: \href{mailto:lee-sangwon@g.ecc.u-tokyo.ac.jp}{lee-sangwon@g.ecc.u-tokyo.ac.jp} (S. Lee).}

		\begin{abstract}
			Model updating of engineering systems inevitably involves handling both aleatory or inherent randomness and epistemic uncertainties or uncertainities arising from a lack of knowledge or information about the system.
			Addressing these uncertainties poses significant challenges, particularly when data and simulations are limited.
			This study proposes a novel latent space-based method for stochastic model updating that leverages limited data to effectively quantify uncertainties in engineering applications.
			By extending the latent space-based approach to multiobservation and multisimulation frameworks, the proposed method circumvents the need for probability estimations at each iteration of MCMC, relying instead on an amortized probabilistic model trained using a variational autoencoder (VAE).
			This method was validated through numerical experiments on a two-degree-of-freedom shear spring model, demonstrating superior efficiency and accuracy compared to existing methods in terms of uncertainty quantification (UQ) metrics, such as Bhattacharyya and Euclidean distances.
			Moreover, the applicability of the method to time-series data was verified using the model calibration problem of the NASA UQ Challenge 2019.
			The results underscore the potential of the latent space-based method in practical engineering applications, providing a robust framework for uncertainty quantification with fewer data requirements, and demonstrating its effectiveness in handling high-dimensional data.
		\end{abstract}

		\begin{keyword}
			Stochastic model updating \sep Latent space \sep Uncertainty quantification \sep Bayesian inference \sep Variational Autoencoder \sep NASA UQ challenge
		\end{keyword}
	\end{frontmatter}
	
	\section{Introduction}
		The model updating of engineering systems based on simulations involves uncertainties from various sources \cite{GoldsteinM2016}.
		These uncertainties can be categorized into two types: aleatory and epistemic.
		Aleatory uncertainties are unavoidable and should be considered in simulation models.
		Epistemic uncertainties, however, are those that can potentially be reduced through the collection of additional data or by refining models \cite{Mullins16}.
		
		Effectively reducing epistemic uncertainties and accurately quantifying aleatory uncertainties necessitate gathering multiple observations and performing multiple simulations to ensure comprehensive data coverage.
		This approach, which can be termed as creating a multiobservation multisimulation scenario, is essential for precise uncertainty quantification \cite{Bi23}.
		Thus, many stochastic model updating methods that explicitly address both epistemic and aleatory uncertainties of model parameters have been proposed \cite{Bi23, Mares06, Khodaparast08, Govers15}.
		These methods leverage statistical and computational techniques to refine the model and improve its accuracy and robustness in the presence of inherent uncertainties.
		
		Among these methods, the Bayesian model updating framework has been used to effectively manage and quantify uncertainties.
		This method systematically updates the model parameters using observed data and merges prior knowledge with new observations to create a robust framework for uncertainty quantification \cite{Beck98}.
		A key aspect of this framework is the likelihood evaluation, which quantifies uncertainty and assesses the probability of observed data based on a specific hypothesis, serving as an essential transition from posterior probabilities.

		However, in many practical scenarios, calculating the full likelihood is either computationally intensive or infeasible.
		To address this challenge, an approximate Bayesian computation (ABC) was designed for situations in which calculating the full likelihood is difficult \cite{Beaumont02, Beaumont09}.
		ABC was initially based on the concept of Monte Carlo rejection sampling \cite{Tavare97, Pritchard99}.
		When applied to stochastic model updating, it is common to define the likelihood based on a Gaussian kernel function grounded in an uncertainty quantification metric, rather than using rejection sampling.
		In the ABC for the stochastic model updating method, defining a comprehensive uncertainty quantification (UQ) metric capable of quantifying the uncertain discrepancy between the observed data and numerical model predictions is crucial.
		Therefore, many UQ metrics based on statistical distances have been used \cite{Bi19a, Bi19b, Kitahara21, Kitahara22}.
		These UQ metrics can efficiently calculate statistical distances using multidimensional histograms of the observation and simulation samples \cite{Lye24}.

		However, when considering real-world applications, creating a multiobservation multisimulation scenario presents two main challenges.
		(1) Increased simulation costs and data dimensionality owing to the sophistication and complexity of engineering systems.
		(2) Data insufficiency from real-world systems.
		Moreover, despite the existence of various metrics, most rely on comparing the probability distributions estimated from histograms, necessitating a substantial amount of data.
		Even for low-dimensional data, hundreds or thousands of observations and simulations are typically required to estimate probability distributions.
		Furthermore, for high-dimensional data, ensuring adequate histogram resolution requires an exponentially increasing number of simulations and observations as data dimensions increase.
		Thus, most existing methods adopt summary statistics to reduce data dimensionality. However, the identification of an appropriate summary statistic in practical contexts is often challenging.
		Given these challenges, a UQ method capable of efficiently evaluating high-dimensional data uncertainty with a limited number of observations and simulations is critical for real-world engineering system applications.

		This study proposes a new method for efficiently quantifying uncertainties in high-dimensional data with fewer observations and simulations than previously suggested methods.
		Specifically, the latent space-based method \cite{Lee24a}, which approximates the likelihood in a nonparametric manner, was extended to a multiobservation, multisimulation scenario.
		This new approach estimates the data density in the latent space using a pretrained probabilistic generative model, specifically, a variational autoencoder (VAE), thereby omitting iterative probability estimation and requiring only a small amount of data.
		Moreover, data-driven dimensionality reduction eliminates the need to manually determine summary statistics, making them robust even as the data dimensions increase.

		The remainder of this study is organized as follows:
		In Section 2, brief explanations of the VAE and the proposed latent space-based method are described.
		In Section 3, the theoretical extension of the latent space-based method for stochastic model updating is discussed, including a comparison with conventional methods that use statistical distances.
		In Section 4, the posterior distributions estimated by the proposed method are compared with the theoretically derived posterior distributions for a two-degree-of-freedom system, demonstrating the superiority of the proposed method through the relationship between the number of observations, simulations, and their accuracy.
		In Section 5, the application of the method to the NASA UQ challenge is discussed in detail, highlighting improvements in predictive accuracy and alignment with the observed data. The effectiveness of this method in handling high-dimensional uncertainty quantification tasks was evaluated, demonstrating its potential for broader applications.

	\section{Theories and methodologies}
		
		\subsection{Bayesian model updating}
			In this study, Bayesian model updating was employed as the underlying framework by incorporating a latent space-based method.
			Bayesian updating was performed by evaluating the conditional probabilities of the parameters based on the observed feature sample \( p\left(\boldsymbol{\uptheta}| \mathbf{x}_\text{obs}\right) \) expressed as
			\begin{equation}
				p\left(\boldsymbol{\uptheta} | \mathbf{x}_\text{obs}\right) = \frac{p\left(\mathbf{x}_\text{obs} | \boldsymbol{\uptheta}\right) p\left(\boldsymbol{\uptheta}\right)}{p\left(\mathbf{x}_\text{obs}\right)}
			\end{equation}
			where \(\boldsymbol{\uptheta}\) represents the parameters of the engineering system model that are the targets for updating. Variable \(\mathbf{x}\) denotes the feature sample representing the behavior of the engineering system, and \(\mathbf{x}_{\text{obs}}\) is the observed sample of \(\mathbf{x}\).
			In this equation:
			\begin{itemize}
				\item \( p\left(\boldsymbol\uptheta | \mathbf{x}_\text{obs}\right) \) represents the posterior distribution, reflecting the updated knowledge based on the observational data;
				\item  \( p\left(\mathbf{x}_\text{obs}|\boldsymbol\uptheta\right) \) denotes the likelihood function of parameters \( \boldsymbol\uptheta \) given  the observed data \( \mathbf{x}_\text{obs} \);
				\item \( p(\boldsymbol\uptheta) \) serves as the prior distribution, representing initial knowledge about \( \boldsymbol\uptheta \);
				\item \( p\left(\mathbf{x}_\text{obs}\right) \) is the normalization constant ensuring the posterior distribution sums to unity.
			\end{itemize}
			
			Given that \( p\left(\mathbf{x}_\text{obs}\right) \) is constant, the posterior distribution can be reformulated as follows:
			\begin{equation}
				\label{bayes1}
				p\left(\boldsymbol\uptheta|\mathbf{x}_\text{obs}\right)\propto
				p\left(\mathbf{x}_\text{obs}|\boldsymbol\uptheta\right) p(\boldsymbol\uptheta)
			\end{equation}
			Assuming that prior distribution \( p(\boldsymbol\uptheta) \) is derived from prior information, only the likelihood \( p\left(\mathbf{x}_\text{obs}|\boldsymbol\uptheta\right) \) should be determined from the observations.
		
		\subsection{Latent space-based likelihood estimation}
			The likelihood $p(\mathbf{x}_{\text{obs}}|\boldsymbol{\uptheta})$ can be marginalized using an arbitrary multidimensional random variable $\mathbf{z} \in \mathcal{Z}$ such that $ p(\mathbf{x}_{\text{obs}}|\mathbf{z},\boldsymbol{\uptheta})=p(\mathbf{x}_{\text{obs}}|\mathbf{z})$ as follows:
			\begin{equation}
				p(\mathbf{x}_{\text{obs}}|\boldsymbol{\uptheta}) =  \int_{\mathcal{Z}} p(\mathbf{x}_{\text{obs}}|\mathbf{z}) p(\mathbf{z}|\boldsymbol{\uptheta}) d\mathbf{z}
			\end{equation}
			By applying Bayes' theorem and considering that \( p(\mathbf{x}_{\text{obs}}) \) is a constant, it can be expressed as
			\begin{equation}
				p(\mathbf{x}_{\text{obs}}|\boldsymbol{\uptheta})  \propto   \int_{\mathcal{Z}} \frac{p(\mathbf{z}|\mathbf{x}_{\text{obs}})p(\mathbf{z}|\boldsymbol{\uptheta})}{p(\mathbf{z})} d\mathbf{z}
			\end{equation}
			Here, using the approximation \( q \) for the probability density function \( p \), the likelihood can be determined as
			\begin{equation}
				\label{eq:lsmu_lik}
				p(\mathbf{x}_{\text{obs}}|\boldsymbol{\uptheta})  \simeq \int_{\mathcal{Z}} \frac{q(\mathbf{z}|\mathbf{x}_{\text{obs}})q(\mathbf{z}|\boldsymbol{\uptheta})}{q(\mathbf{z})} d\mathbf{z}
			\end{equation}
			In this study, the encoder of a pre-trained VAE \cite{VAE} was used as \( q \) in Eq. (\ref{eq:lsmu_lik}) to calculate the data-driven likelihood \cite{Lee23}.
		
		\subsection{Overview of VAE}
			The VAE, originally proposed by Kingma and Welling \cite{VAE}, is a generative model that includes both generative and inferential models.
			This is a prominent example of a generative model, which is a type of unsupervised learning method.
			The VAE is specifically designed to replicate the training data by learning an underlying latent representation inferred from the observed data and capturing the hidden structure within the dataset.

			\begin{figure}[t]
				\centering
				\includegraphics[width=0.8\linewidth]{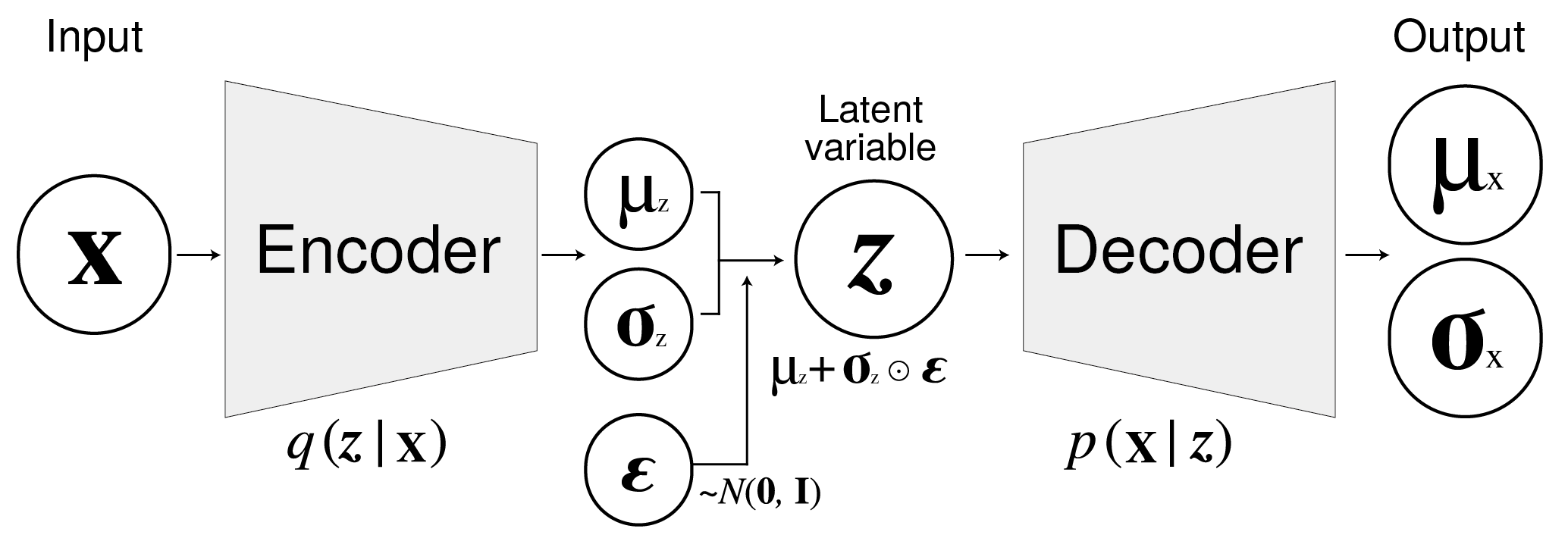}
				\caption{Graphical model of VAE}
				\label{fig:vae}
			\end{figure}
			
			Fig. \(\ref{fig:vae}\) shows the graphical representation of a VAE.
			In the VAE, data \(\mathbf{x}\) can be reconstructed or transformed from an \(N_z\)-dimensional random variable \(\mathbf{z} \in \mathbb{R}^{N_z}\) using a generative model known as the decoder.
			The variable \(\mathbf{z}\) is referred to as a latent variable. The dimension of \(\mathbf{z}\), \(N_z\), is typically smaller than that of \(\mathbf{x}\).
	
			The inverse of the generative model, which corresponds to the posterior distribution of \(\mathbf{z}\), \(p(\mathbf{z}|\mathbf{x})\), represents the inference of the latent variable \(\mathbf{z}\) hidden behind the data \(\mathbf{x}\).
			The encoder approximates the posterior \(p(\mathbf{z}|\mathbf{x})\) through a model denoted as \(q(\mathbf{z}|\mathbf{x})\).
			The decoder and the encoder of the VAE are modeled using neural networks.

			The training of the VAE aimed to better represent the training dataset with minimal loss of information.
			The loss function $Loss(\mathbf{x})$ for training the VAE included two components: the reconstruction and regularization terms:
			\begin{equation}
				\label{eq:ob_func}
				Loss(\mathbf{x}) = -\mathbb{E}_{q(\mathbf{z}|\mathbf{x})} \left[ \log p(\mathbf{x}|\mathbf{z}) \right] + D_{KL} \left[ q(\mathbf{z}|\mathbf{x}) \parallel p(\mathbf{z}) \right]
			\end{equation}
			where $\mathbb{E}_{q(\mathbf{z}|\mathbf{x})} [\cdot]$ is the expectation operator on $q(\mathbf{z}|\mathbf{x})$, and $D_{KL} \left[ q(\mathbf{z}|\mathbf{x}) \parallel p(\mathbf{z}) \right]$ is the Kullback–Leibler distance between $q(\mathbf{z}|\mathbf{x})$ and $p(\mathbf{z})$.
			The first and second terms on the right side of Eq. (\ref{eq:ob_func}) correspond to the reconstruction terms of the decoder and the regularization term, respectively.
			This minimization ensures that the probability distribution of $\mathbf{z}$, where each element is independent of the others, closely approximates the standard normal distribution.

	\section{Latent space-based stochastic model updating}
		
		\subsection{Stochastic model updating in a multiobservation multisimulation scenario}
			Consider the aleatory uncertainty of a parameter obtained through an arbitrary probability distribution \( f_\theta \) and its hyperparameter  \( \boldsymbol{\upphi} \) as shown in the following expression:
			\begin{equation}
				\boldsymbol{\uptheta} \sim f_\theta(\boldsymbol{\upphi})
			\end{equation}
			Here, we aim to evaluate the uncertainty associated with the unknown \( \boldsymbol{\upphi} \).
			When matrix \( \mathbf{X}_{\text{obs}} \) is obtained, the likelihood of \( \boldsymbol{\upphi} \), $p(\mathbf{X}_{\text{obs}}|\boldsymbol{\upphi})$, can be expressed as follows:
			\begin{equation}
				p(\mathbf{X}_{\text{obs}}|\boldsymbol{\upphi})= \prod_{i=1}^{N_{obs}} p(\mathbf{x}_{\text{obs}}^{(i)}|\boldsymbol{\upphi})
			\end{equation}
			where
			\begin{equation}
				\mathbf{X}_{\text{obs}} = [\mathbf{x}_{\text{obs}}^{(1)},\mathbf{x}_{\text{obs}}^{(2)}, \dots, \mathbf{x}_{\text{obs}}^{(N_{obs})}]
			\end{equation}
			represents a matrix of observations.
			Using Eq. (\ref{eq:lsmu_lik}), the equation can be transformed according to the latent variable $\mathbf{z}$ as follows:
			\begin{equation}
				p(\mathbf{X}_{\text{obs}}|\boldsymbol{\upphi}) \propto \ \prod_{i=1}^{N_{obs}} \int_{\mathcal{Z}} \frac{p(\mathbf{z}|\boldsymbol{\upphi}) \ p(\mathbf{z}| \mathbf{x}_{\text{obs}}^{(i)})}{p(\mathbf{z})} \ d\mathbf{z}
			\end{equation}
			Here,  \( p(\mathbf{z}|\boldsymbol{\upphi}) \) is evaluated by marginalizing  \( p(\mathbf{z},\boldsymbol{\uptheta} |\boldsymbol{\upphi})  \)  over parameter \(\boldsymbol{\uptheta}\). Thus,
			\begin{equation}
				p(\mathbf{X}_{\text{obs}}|\boldsymbol{\upphi}) \propto \ \prod_{i=1}^{N_{obs}} \int_{\mathcal{Z}} \frac{\left( \int_{\mathcal{\theta}}  p(\mathbf{z} | \boldsymbol{\uptheta}) p(\boldsymbol{\uptheta}|\boldsymbol{\upphi})  d\boldsymbol{\uptheta} \right) \ p(\mathbf{z}| \mathbf{x}_{\text{obs}}^{(i)})}{p(\mathbf{z})} \ d\mathbf{z}
			\end{equation}
			By integrating over \(\boldsymbol{\uptheta}\) using Monte Carlo methods, we can approximate the likelihood as follows:
			\begin{equation}
				p(\mathbf{X}_{\text{obs}}|\boldsymbol{\upphi})\simeq  \ \prod_{i=1}^{N_{obs}} \sum_{j=1}^{N_{sim}} \int_{\mathcal{Z}} \frac{ p(\mathbf{z}| \boldsymbol{\uptheta}_j^{*})  \ p(\mathbf{z}| \mathbf{x}_{\text{obs}}^{(i)})}{p(\mathbf{z})} \ d\mathbf{z}
			\end{equation}
			where \( \boldsymbol{\uptheta}_j^{*} \) is the \( j \)th sample randomly obtained from \( p(\boldsymbol{\uptheta}|\boldsymbol{\upphi}) \) and \( N_{sim} \) represents a sufficient number of samples for Monte Carlo integration.
			Finally, the likelihood can be approximated using a data-driven probabilistic model $q$ instead of an intractable $p$.
			\begin{equation}
				\label{eq:smu_lik}
				p(\mathbf{X}_{\text{obs}}|\boldsymbol{\upphi}) \simeq \ \prod_{i=1}^{N_{obs}} \sum_{j=1}^{N_{sim}} \int_{\mathcal{Z}} \frac{q(\mathbf{z}| \boldsymbol{\uptheta}_j^{*}) \ q(\mathbf{z}| \mathbf{x}_{\text{obs}}^{(i)})}{q(\mathbf{z})} \ d\mathbf{z}
			\end{equation}
			Since \( q(\cdot) \) derived from the VAE follows a Gaussian distribution, Eq. (\ref{eq:smu_lik}) can be analytically calculated using the detailed calculation provided in \ref{sec:app0}.
			In Eq. (\ref{eq:smu_lik}), the proposed method assesses the degree of similarity between the data and simulation distributions by calculating the weighted inner product of the Gaussian mixtures $\sum_{j=1}^{N_{sim}}q(\mathbf{z}| \boldsymbol{\uptheta}_j^{*})$  and the Gaussian $q(\mathbf{z}| \mathbf{x}_{\text{obs}}^{(i)})$.
			If \( q \) approximates \( p \) well with a sufficient \( N_{sim} \), then evaluating the likelihood of being identical to the theoretical probability is feasible.
			
			The pseudocode for calculating the extended likelihood and detailing the primary steps and repetitive processes is presented in Algorithm \ref{al:2}.
			\begin{algorithm}
				\caption{Latent Space-based Likelihood Calculation for Stochastic Model Updating} \label{al:2}
				\begin{algorithmic}[1]
					\State \textbf{Input:} Hyperparameters $\boldsymbol{\upphi}$, Observed data $\mathbf{X}_{\text{obs}} = [\mathbf{x}_{\text{obs}}^{(1)}, \mathbf{x}_{\text{obs}}^{(2)}, \dots, \mathbf{x}_{\text{obs}}^{(N_{obs})}]$, Encoding neural network $Enc(\cdot)$, Simulator $h(\cdot)$, Number of simulations $N_{sim}$
					\State \textbf{Output:} Likelihood $L$
					\For{each observed data point $i = 1$ to $N_{obs}$}
						\State Encode observed data using the neural network: $q(\mathbf{z}|\mathbf{x}_{\text{obs}}^{(i)})= Enc(\mathbf{x}_{\text{obs}}^{(i)})$
						\For{each simulation $j = 1$ to $N_{\text{sim}}$}
							\If{$j = 1$}
								\State Sample $\boldsymbol{\uptheta}_j^{\ast} \sim p(\boldsymbol{\uptheta}|\boldsymbol{\upphi})$
								\State Simulate data $\mathbf{x}_{\text{sim}}^{(j)} = h(\boldsymbol{\uptheta}_j^{\ast})$
								\State Encode simulated data using the neural network: $q(\mathbf{z}|\mathbf{x}_{\text{sim}}^{(j)})= Enc(\mathbf{x}_{\text{sim}}^{(j)})$
							\EndIf
							\State Compute the  likelihood : $l_{i,j} = \mathcal{L}_{\text{cal}}(q(\mathbf{z}|\mathbf{x}_{\text{obs}}^{(i)}), q(\mathbf{z}|\mathbf{x}_{\text{sim}}^{(j)}))$
							\State \Comment{$\mathcal{L}_{\text{cal}}$ computes the likelihood as described in Eq. (\ref{eq:lsmu_lik})}
						\EndFor
					\EndFor
					\State Compute the total likelihood: $L = \prod_{i=1}^{N_{obs}}\sum_{j=1}^{N_{sim}}l_{i,j}$
				\end{algorithmic}
			\end{algorithm}

		\subsection{Comparison of binning algorithm and latent space-based method}
\begin{figure}[H]
				\begin{tabular}{cc}
					\begin{minipage}[t]{0.48\hsize}
						\centering
						\includegraphics[width=0.98\linewidth]{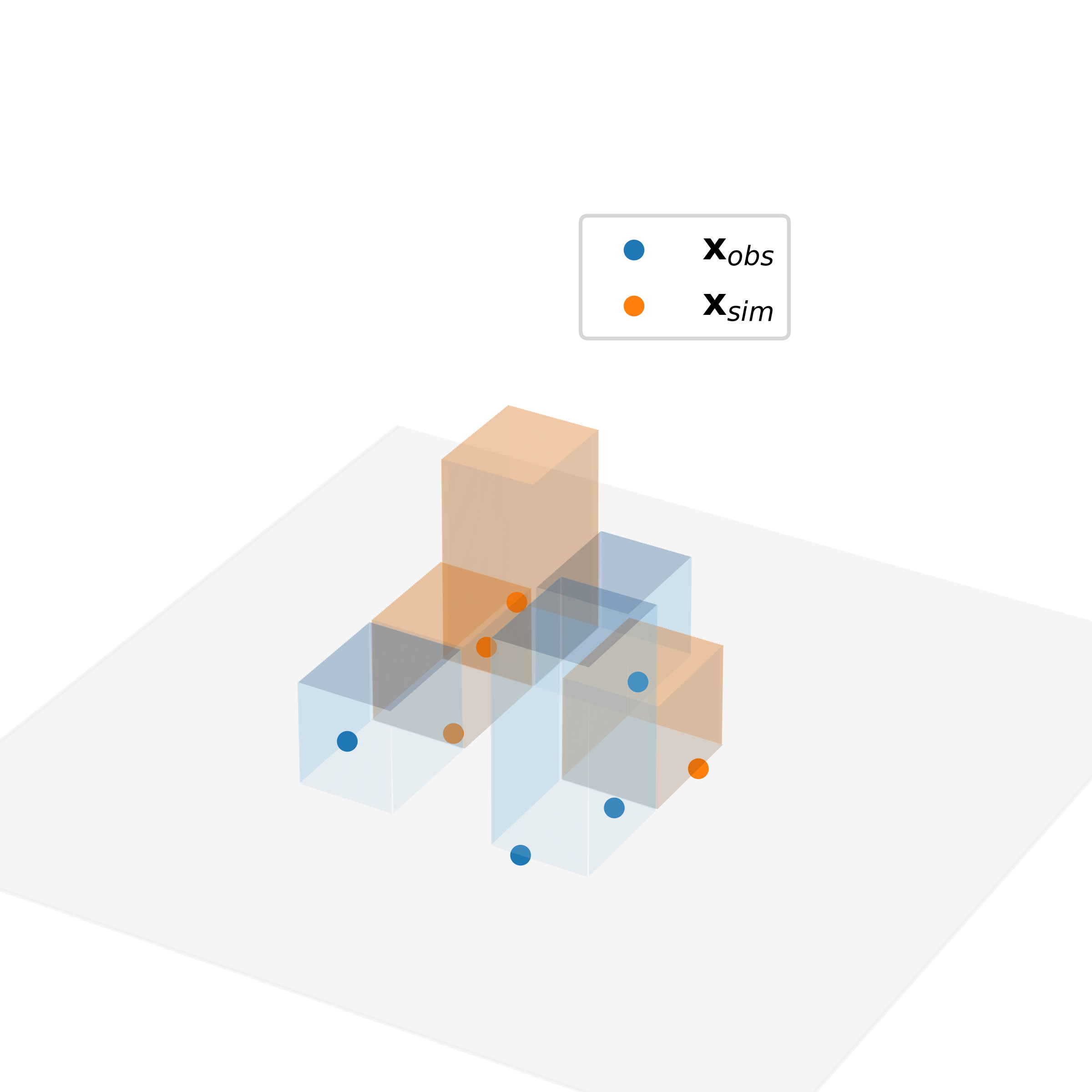}
						\subcaption{Binning algorithm}
					\end{minipage}
					\begin{minipage}[t]{0.48\hsize}
						\centering
						\includegraphics[width=0.98\linewidth]{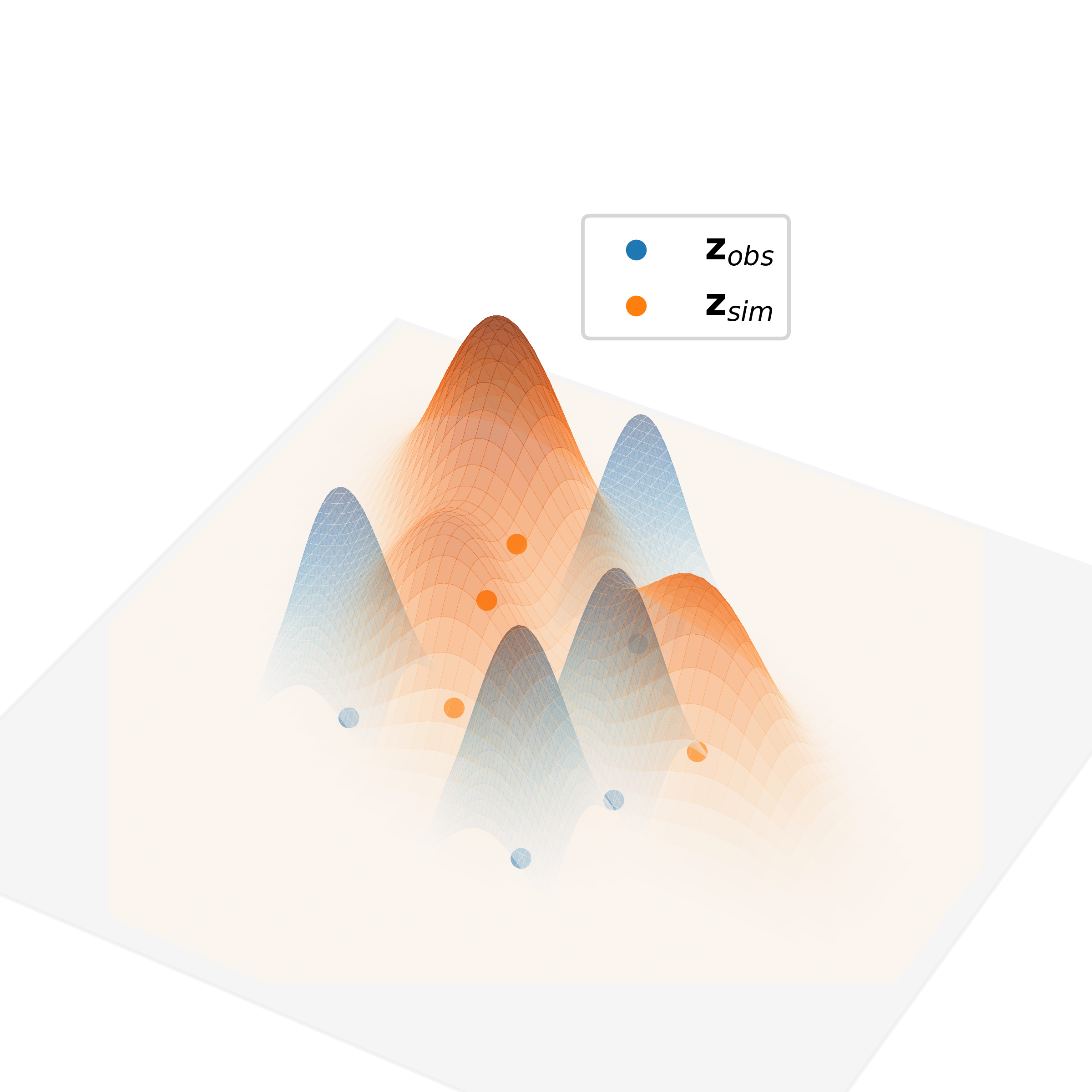}
						\subcaption{Latent space-based method}
					\end{minipage}\\
					\\
				\end{tabular}
				
				\caption{Schematics of probability distribution estimation}\label{fig:comparing_method}
			\end{figure}

			Conventional distance-based methods such as those that use the Bhattacharyya distance involve a process called binning to estimate the probability mass functions (PMF) of the observed and simulated data distributions.
			Fig. \ref{fig:comparing_method} is a schematic of the probability distribution estimation using the binning algorithm and latent space-based method.
			Fig. \ref{fig:comparing_method} (a) illustrates the binning approach, where \( \mathbf{x}_{\text{obs}} \) represents observed data and \( \mathbf{x}_{\text{sim}} \) represents simulated data.
			The data points were binned into discrete intervals, and the height of each bin indicates the frequency of the data points within that interval.
			\noindent Fig. \ref{fig:comparing_method} (b) illustrates the latent space-based method, with \( \mathbf{z}_{\text{obs}} \) representing the latent space projection of observed data and \( \mathbf{z}_{\text{sim}} \) representing the latent space projection of simulated data.
			The continuous surface indicates the probability density function (PDF) modeled by the VAE, assuming an independent Gaussian distribution for \( \mathbf{z}_{\text{obs}} \) and a Gaussian mixture distribution for \( \mathbf{z}_{\text{sim}} \).
			
			Traditional binning algorithms discretize continuous data into finite intervals or bins, enabling the calculation of statistical distances. This approach has several strengths and limitations:
			\begin{itemize}
				\item \textbf{Discrete PMF:} Binning algorithms enable quick calculation of the PMF, rendering them advantageous when handling large numbers of observations (\( N_{obs} \)) and simulations (\( N_{sim} \)).
				\item \textbf{Extensive data requirements:} Reliable PMF estimates require a significant amount of data. The required number of observations (\( N_{obs} \)) and simulations (\( N_{sim} \)) can become prohibitively large, as the complexity of the data increases the need for more detailed binning to capture the distribution accurately.
				\item \textbf{Iteration-specific estimation:} During the MCMC process, the PMF should be estimated at every iteration. Although the computation of PMF in each iteration is relatively fast, the computational burden increases with the dimensionality of the data.
				\item \textbf{Hyperparameter selection:} The number of bins (\( N_{bin} \)) and the width factor should be carefully chosen, with \( N_{bin} \) directly influencing the resolution and accuracy of the PMF, while the width factor primarily affects the subsequent calculation of likelihood from the PMF.
			\end{itemize}
			
			Contrarily, the latent space-based method leverages a VAE to model data distribution in a continuous latent space, offering distinct advantages and considerations.
			\begin{itemize}
				\item \textbf{Continuous PDF:} This method models the data distribution as a continuous PDF, providing high-resolution probabilistic information without requiring extensive data. This approach is more efficient and accurate for high-dimensional data.
				\item \textbf{Reduced data requirements:} The latent space representation significantly lowers the amount of data required for reliable uncertainty quantification. The VAE, trained during the pre-processing phase, captures essential features of the data distribution, enabling effective uncertainty evaluation with fewer observations and simulations.
				\item \textbf{Pretrained model utilization:} The use of a pretrained VAE model enables efficient and accurate probability estimation without the need to recalculate distributions at every iteration, making the method particularly  powerful in high-dimensional contexts.
				\item \textbf{Hyperparameter selection:} In the latent space-based method, key hyperparameters such as the architecture of the VAE (e.g., the number of layers and nodes) and the choice of latent space distribution should be selected. These decisions impact the model’s ability to capture the underlying data distribution accurately.
			\end{itemize}
	
			In summary, both methods are suitable. Binning algorithms are particularly advantageous when there is a large amount of data owing to their quick PMF calculations and straightforward implementation.
			However, as the dimensionality of the data increases, the data requirements for these methods can become overwhelming and the computational burden of iteration-specific estimations increases.
			Conversely, the latent space-based method excels with smaller valuable datasets and high-dimensional data, providing detailed probabilistic information with fewer data points and simulations, and offering a scalable solution for complex data distributions.
			This study emphasizes the benefits of the latent space-based method while acknowledging the efficacy of traditional binning methods in appropriate scenarios.

	\section{Verification with theoretical likelihood calculation}\label{sec:2dof}
		This experiment demonstrates that the proposed method can achieve stochastic model updating using a limited number of observations and simulations.
		Using a problem in which the likelihood can be calculated theoretically, we evaluated the performance of the proposed method and compared it with traditional methods, demonstrating its efficiency and accuracy.
		
		\subsection{Problem description}
\begin{figure}[b]
				\centering
				\includegraphics[width=0.3\linewidth]{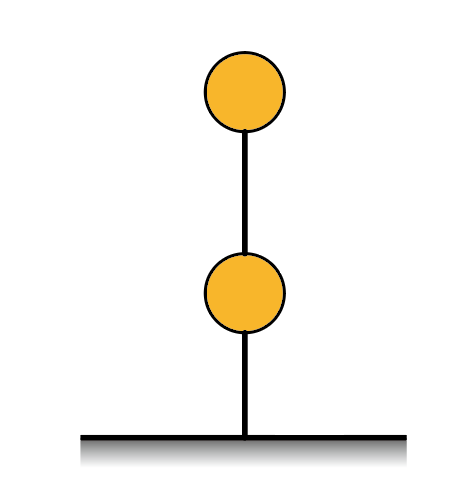}
				\caption{Two degree-of-freedom shear spring system}\label{fig:model}
			\end{figure}

			The proposed method was demonstrated using a two-degree-of-freedom (DOF) shear spring model (Fig. \ref{fig:model}), a model widely used in literature for validating model updating methods \cite{Beck02,Betz18,Wang20,Kitahara23}.
			
			The first- and second-story masses were considered to be deterministic and known values with $m_1 = 16.5 $ tons and $m_2 = 16.1 $ tons.
			The first- and second-story stiffnesses were parameterized as $k_1= \bar{k} \theta_1$ and $k_2 = \bar{k} \theta_2$, respectively, where $\boldsymbol{\uptheta} = [\theta_1, \theta_2]^{\top}$ represents the parameters to be  updated and  $\bar{k} = 29.7$ MN/m is the nominal value of the stiffnesses.
			Furthermore, it was assumed that parameters $\boldsymbol{\uptheta}$ incorporate the following  aleatory uncertainty through hyperparameters $\boldsymbol{\mu} = [\mu_1, \mu_2]$ and $\sigma$:
			\begin{equation}
				\boldsymbol{\uptheta} \sim \mathcal{N}(\boldsymbol{\mu},\sigma^2\mathbb{I})
			\end{equation}
			 where $\mathbb{I} \in \mathbb{R}^2$ denotes an identity matrix.
			The intervals of means $\mu_1$, $\mu_2$ and common standard deviations $\sigma$ associated with the parameters, as well as the target values for the hyperparameters, are detailed in Table \ref{tab:parameter}.

			The following describes observation $\mathbf{x}_{\text{obs}}$:
			\begin{equation}
				\mathbf{x}_{\text{obs}} = [ f_1, f_2, \beta_1\phi_{1, 1}, \beta_1\phi_{1, 2}, \beta_2\phi_{2, 1}, \beta_2\phi_{2, 2}]^{\top} \in \mathbb{R}^6
			\end{equation}
			where  $f_1$ and  $f_2$ are the first and second natural frequencies, respectively, and $\beta_j\phi_{j, k}$ denotes the $k$th component of the $j$th participation function.
			Assuming that $\mathbf{x}_{\text{obs}}$ is obtained $N_{obs}$ times, hyperparameters $\mu_1$, $\mu_2$, and $\sigma$ can be estimated.
			In addition, it was assumed as prior knowledge that the true values of the hyperparameters lie within the ranges listed in Table \ref{tab:parameter}.

			In this numerical experiment, MCMC was used to compare the efficiency of the proposed method with those of the Euclidean distance-based and Bhattacharyya distance-based methods.
			Each method's samples derived from their respective likelihoods were compared with the samples derived from the theoretically calculated likelihood (hereafter referred to as the theoretical likelihood), as detailed in \ref{sec:app1}.
			
		\subsection{Training of VAE}
			The dataset used to train the VAE was created by generating uniform random numbers of hyperparameters within the range specified in TABLE~\ref{tab:parameter}.
			Only parameters $\theta_i$ that fell within the range of $[0.1, 2]$ were included in the dataset.
			The features $\mathbf{x}$ were then derived from the eigenvalue analysis of these filtered parameters.
			VAEs are known to encounter instability issues when trained on low-dimensional data \cite{Dai20}.
			Therefore, in this validation, we used the frequency response function to stabilize the training by transforming it into higher dimension data.
			The frequency response function used is as follows:
			\begin{equation}
				H_k(f(k)) = \sum_{j=1}^2 \left| \beta_j\phi_{j,k}\left( \frac{{f(k)}^2}{{f_{j}}^2 - {f(k)}^2 + 2 i h_j f_{j} f(k)} +1 \right) \right|
			\end{equation}
			Here, $h_j$ is the $j$th damping constant and both were set to 0.05. The high-dimensional features derived from this function were used for training with $f(k) = 0.02 \times (k-1)$ for $k = 1, 2, \dots, 512$.
			
			\begin{table}[t]
				\caption{Parameter intervals and ground truth values}\label{tab:parameter}
				\centering
				\begin{tabular}{cccc}
					\hline
					parameter & Distribution & Uncertainty characteristics               & Ground truth                    \\
					\hline
					$\theta_1 $    & Gaussian     & $\mu_1 \in [0.1, 2]$, $\sigma \in [0, 0.6] $ & $\mu_1 = 0.49 $, $\sigma = 0.1$ \\
					$\theta_2 $    & Gaussian     & $\mu_2 \in [0.1, 2]$, $\sigma \in [0, 0.6] $ & $\mu_2 = 0.92 $, $\sigma = 0.1$ \\
					\hline
				\end{tabular}
			\end{table}
			
			The dataset was structured as a tensor with dimensions (100,000, 2, 1, 512), where the values in parentheses denote the number of data points, number of channels (representing the frequency response function at each floor), width, and length (corresponding to the frequency points of the frequency response function).
	
\begin{figure}[t]
				\centering
				\includegraphics[width=0.97\linewidth]{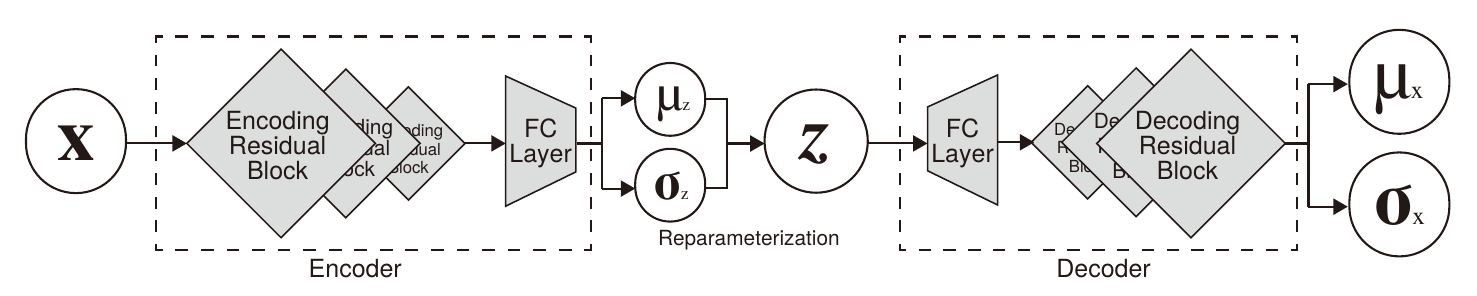}
				\caption{Structure of VAE} \label{fig:vaenet}
			\end{figure}

			The architecture of the VAE network used to update the model is illustrated in Fig. \ref{fig:vaenet}. The VAE consisted of an encoder that converts data $\mathbf{x}$ into $\boldsymbol{\upmu}_\mathbf{z}$ and $\boldsymbol{\upsigma}_\mathbf{z}$ of latent variable $\mathbf{z}$, and a decoder that reconstructs the latent variable $\mathbf{z}$ back into $\boldsymbol{\upmu}_\mathbf{x}$ and $\boldsymbol{\upsigma}_\mathbf{x}$ of data $\mathbf{x}$.
			The encoder reduces the dimensionality of the input using residual blocks \cite{resnet} that include convolutional neural network (CNN)  and fully connected (FC) layers, with the frequency response function as the input.
			The residual blocks of the encoder double the number of channels during downsampling, thereby reducing the data dimensions by half.
			Conversely, the decoder is structured as the symmetrical counterpart of the encoder, employing residual blocks that halve the number of channels during upsampling, thereby doubling the data dimensions.
			In this study, we set the dimensionality of the latent variable $\boldsymbol{z}$ to three , which slightly exceeds the number of parameters.
			This choice is based on the assumption that the degrees of freedom of data variability are equivalent to the number of parameters.
			
			\begin{figure}[t]
				\centering
				\includegraphics[width=0.7\linewidth]{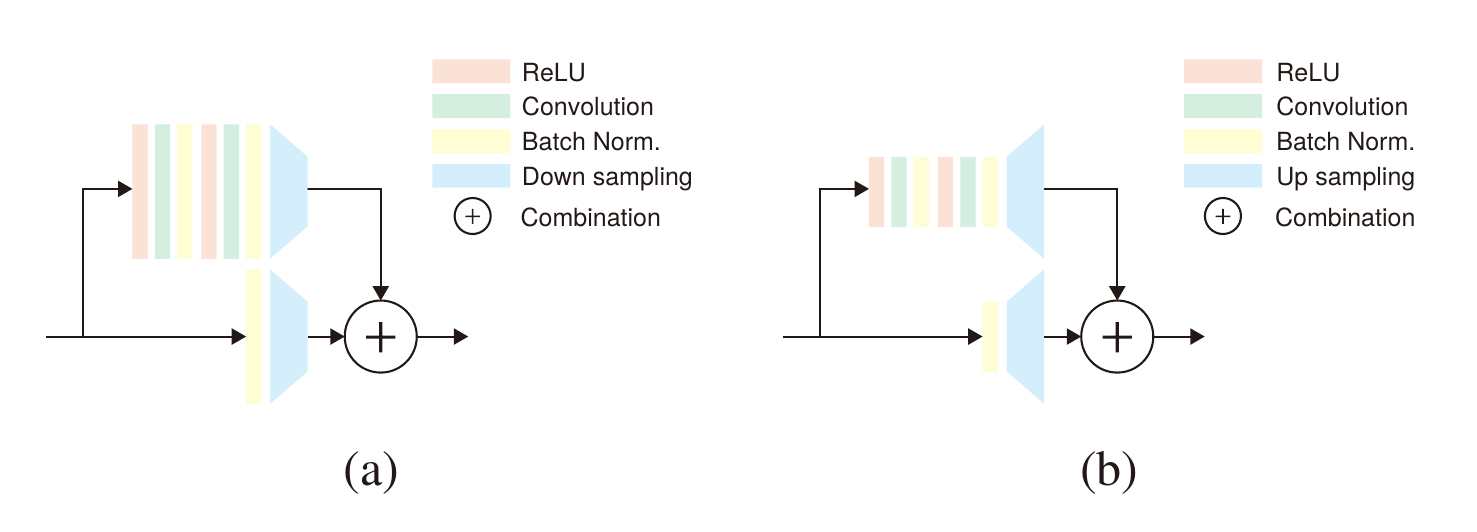}
				\caption{Residual blocks} \label{fig:resblock}
			\end{figure}
		
		\subsection{Sampling with MCMC}
			Stochastic model updating was performed using the proposed method, Bhattacharyya distance-based method, and Euclidean distance-based method.
			To evaluate the  posterior probability, MCMC sampling was conducted using the replica exchange Monte Carlo method \cite{Swendsen86,remc96} with a non-informative prior distribution, generating each replica sample using the Metropolis–Hastings method \cite{metropolis53,Hastings70}.
			
			For the initial values, the values closest to the average of the latent variables $\mathbf{z}$ of the observed samples within the validation dataset were selected for the proposed method, whereas for distance-based methods, the values closest to the average of the observed samples were chosen.
			The exchange was repeated 1,000 times for every 100 samples to obtain 100,000 samples.
			The first 10,000 samples were considered burn-in and discarded, and thinning was performed by selecting every 30th sample, resulting in 3,000 samples. For validation, MCMC sampling was performed using the theoretically derived likelihood.

		\subsection{Results of stochastic model updating}
			To evaluate the performance of each method in likelihood estimation, the  Jensen–Shannon (JS) divergence was calculated between the probability distributions obtained by each method and those derived from the theoretical likelihood.
			In the Euclidean and Bhattacharyya distance-based methods, as summarized in Table \ref{tab:setparameter}, different parameter settings were used to verify three and nine cases, respectively.
			
			\begin{table}[t]
				\caption{Analysis parameter}\label{tab:setparameter}
				\centering
				\begin{tabular}{ccc}
					\hline
					& Width factor $\epsilon$ & Number of bins $N_{bin}$ \\
					\hline
					Euclidean Distance     & 0.1, 0.01, 0.001        & -                        \\
					\hline
					Bhattacharyya Distance & 0.1, 0.01, 0.001          & 3, 5, 7                  \\
					\hline
				\end{tabular}
			\end{table}
			
\begin{figure}[t]
				\begin{tabular}{c}
					\begin{minipage}[t]{0.48\hsize}
						\centering
						\includegraphics[width=0.98\linewidth]{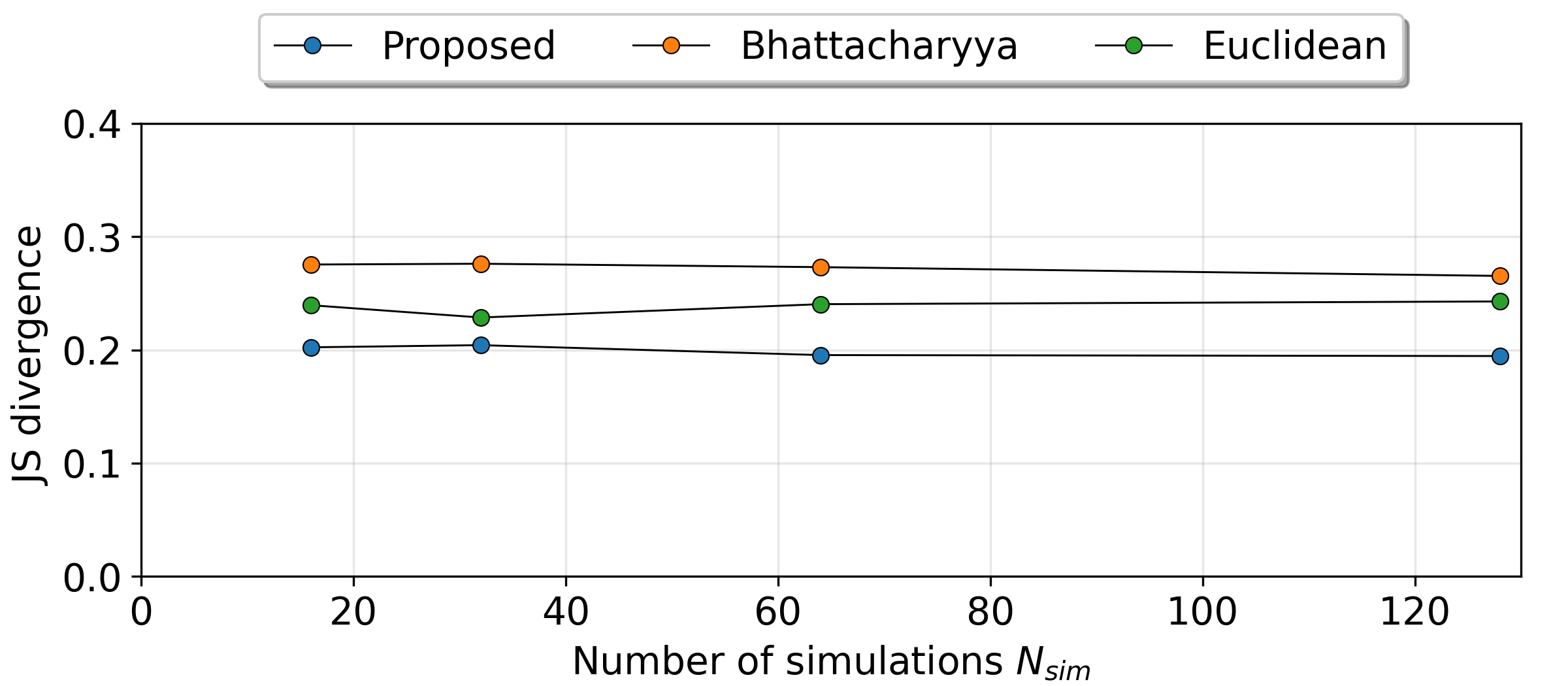}
						\subcaption{$N_{obs}=4$}
					\end{minipage}
					\begin{minipage}[t]{0.48\hsize}
						\centering
						\includegraphics[width=0.98\linewidth]{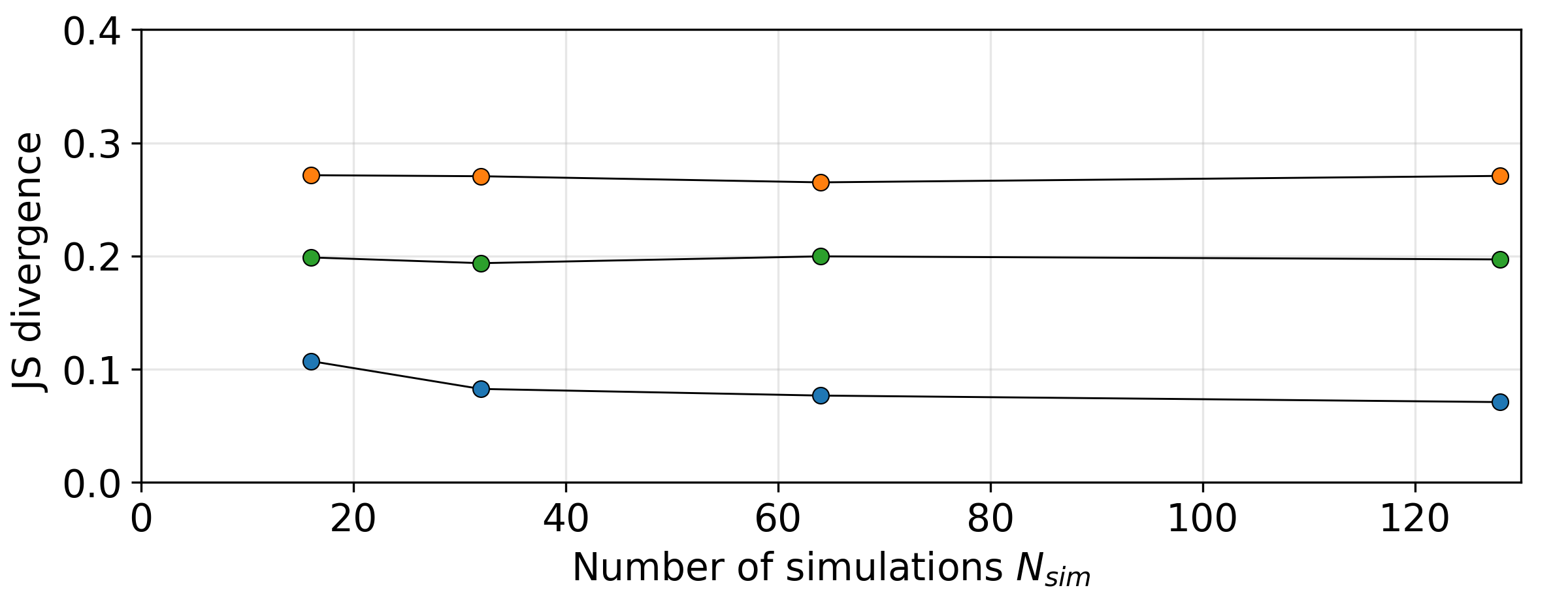}
						\subcaption{$N_{obs}=8$}
					\end{minipage} \\\\
					\begin{minipage}[t]{0.48\hsize}
						\centering
						\includegraphics[width=0.98\linewidth]{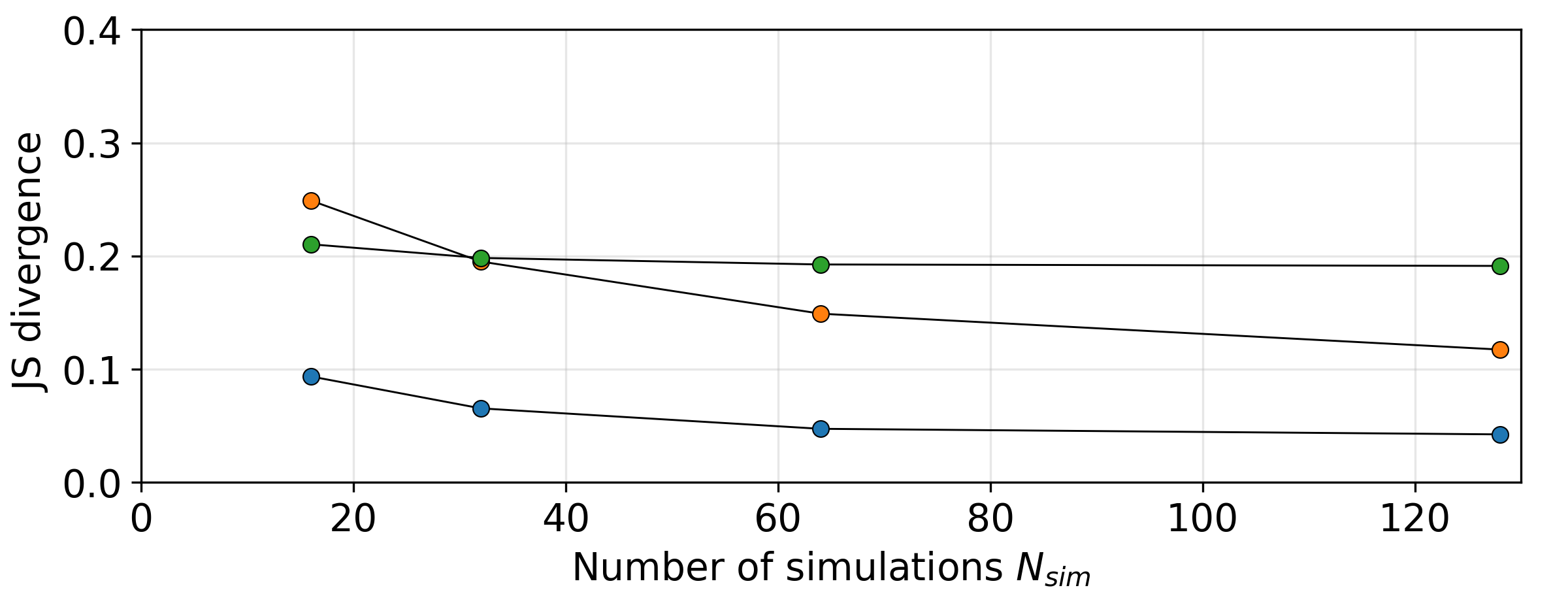}
						\subcaption{$N_{obs}=16$}
					\end{minipage}
					\begin{minipage}[t]{0.48\hsize}
						\centering
						\includegraphics[width=0.98\linewidth]{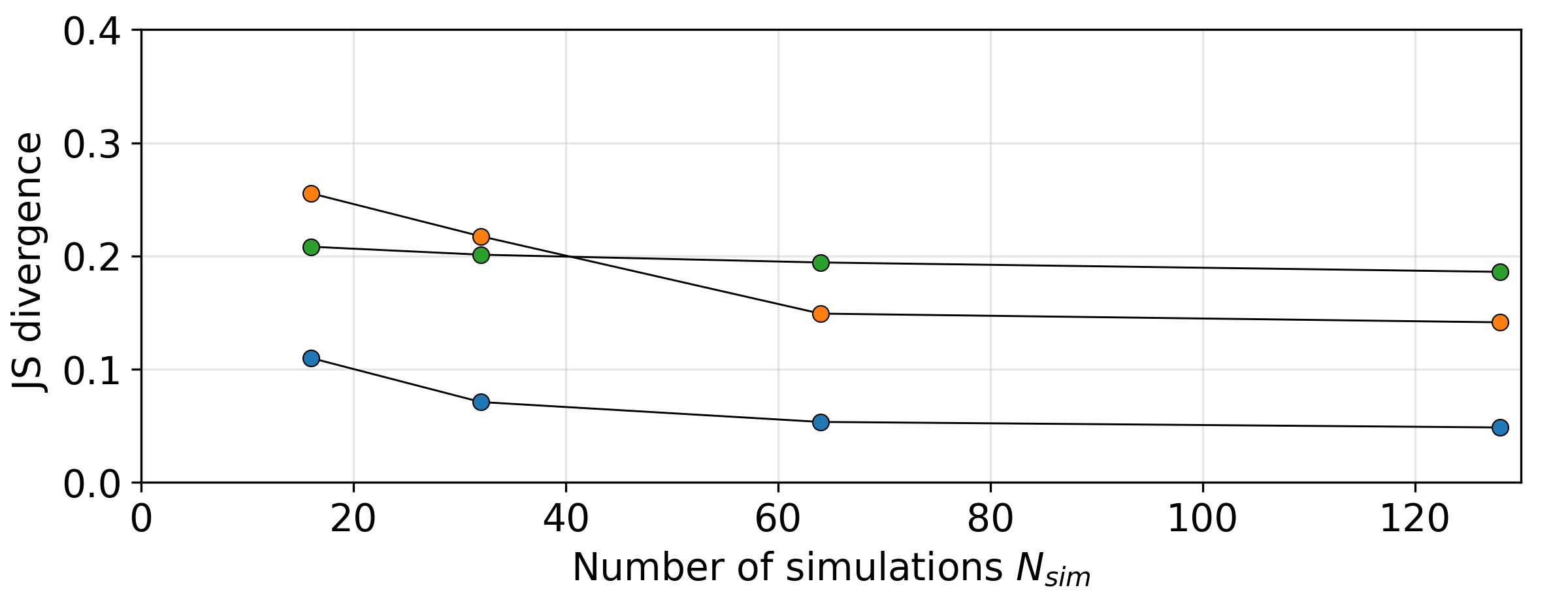}
						\subcaption{$N_{obs}=32$}
					\end{minipage}\\\\
					\begin{minipage}[t]{0.48\hsize}
						\centering
						\includegraphics[width=0.98\linewidth]{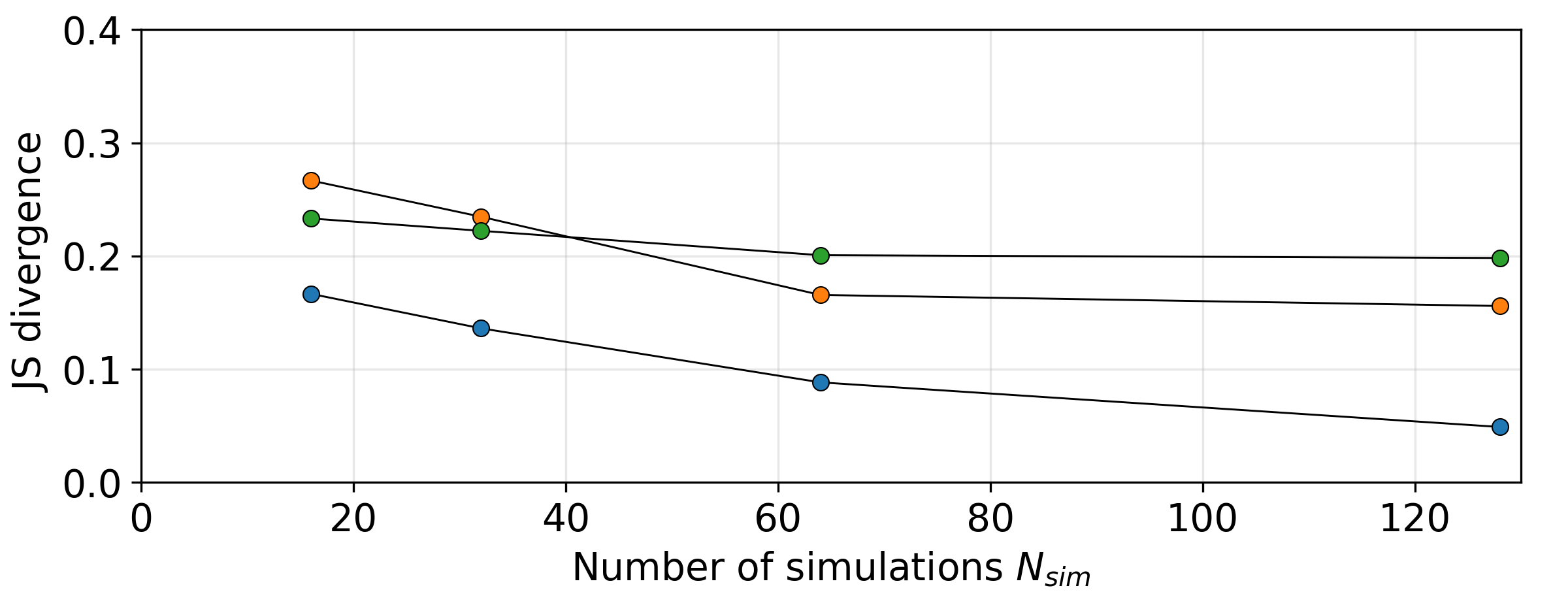}
						\subcaption{$N_{obs}=64$}
					\end{minipage}
					\begin{minipage}[t]{0.48\hsize}
						\centering
						\includegraphics[width=0.98\linewidth]{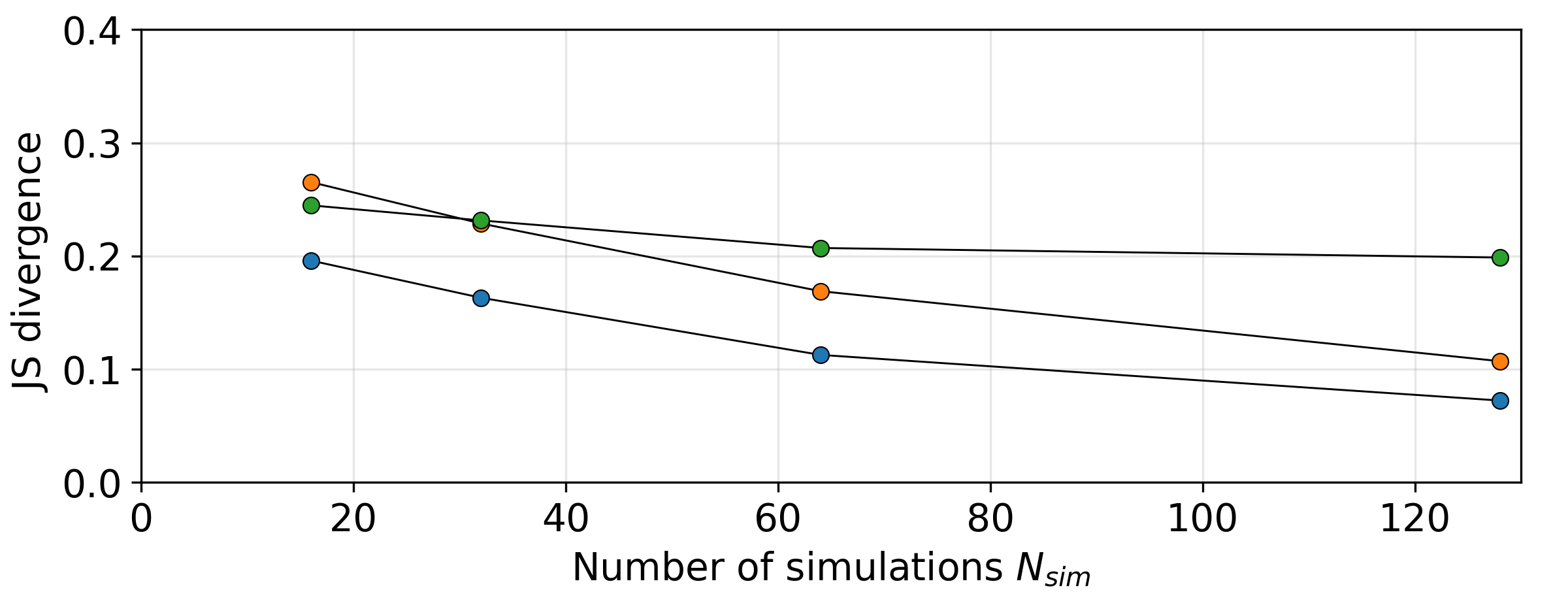}
						\subcaption{$N_{obs}=128$}
					\end{minipage} \\
				\end{tabular}
				\caption{Distance to theoretical likelihood} \label{fig:JSD}
			\end{figure}

			Subsequently, the JS divergence from the samples was computed based on theoretical likelihood.
			The analysis parameter ranges were set according to Table \ref{tab:parameter} and the probability distributions were estimated using multidimensional histograms with 100 bins per dimension.
			The JS divergences for different numbers of observations $N_{obs}$ and simulations $N_{sim}$ are shown in Fig. \ref{fig:JSD}.
			Here, only the results from the  Euclidean distance-based and Bhattacharyya distance-based methods, with the parameter settings listed in Table \ref{tab:setparameter} that resulted in the smallest JS divergence, are presented.

			For all methods, larger \( N_{sim} \) resulted in a smaller JS divergence, demonstrating the benefit of increased simulation counts in improving the likelihood estimation accuracy.
			The proposed method demonstrates a relatively smaller JS divergence across all \( N_{obs} \) and \( N_{sim} \) values, suggesting a higher efficiency, particularly for time-consuming simulations, owing to the fact that fewer simulations were required.
			Notably,  for cases with \( N_{obs} \leq 32 \), \( N_{sim} = 16 \) for the proposed method resulted in a smaller JS divergence than \( N_{sim} = 128 \) for the traditional methods, underscoring its superior performance.
			
			For the Bhattacharyya distance-based method, when \( N_{obs} \) was small (4 and 8), the JS divergence was greater than that for the Euclidean distance-based method. However, for \( N_{obs} \geq 32 \), the JS divergence was smaller than that of the Euclidean distance-based method. Moreover, with sufficient observations and simulations (e.g., \( N_{obs} = 128 \), \( N_{sim} = 128 \)), the Bhattacharyya distance-based method demonstrated a performance comparable to that of the proposed method.

			For both the proposed and  Bhattacharyya distance-based methods, when \( N_{obs} \) was small, increasing \( N_{sim} \) did not significantly affect the JS divergence.
			However, as \( N_{obs} \) increased, the JS divergence decreased more noticeably for higher \( N_{sim} \) values.
			For the proposed method, the JS divergence continued to decrease up to approximately four times the \( N_{obs} \) value for \( N_{sim} \); beyond this point, no further reduction in the JS divergence was observed.

			\begin{figure}[t]
				\begin{tabular}{c}
					\begin{minipage}[t]{0.48\hsize}
						\centering
						\includegraphics[width=0.98\linewidth]{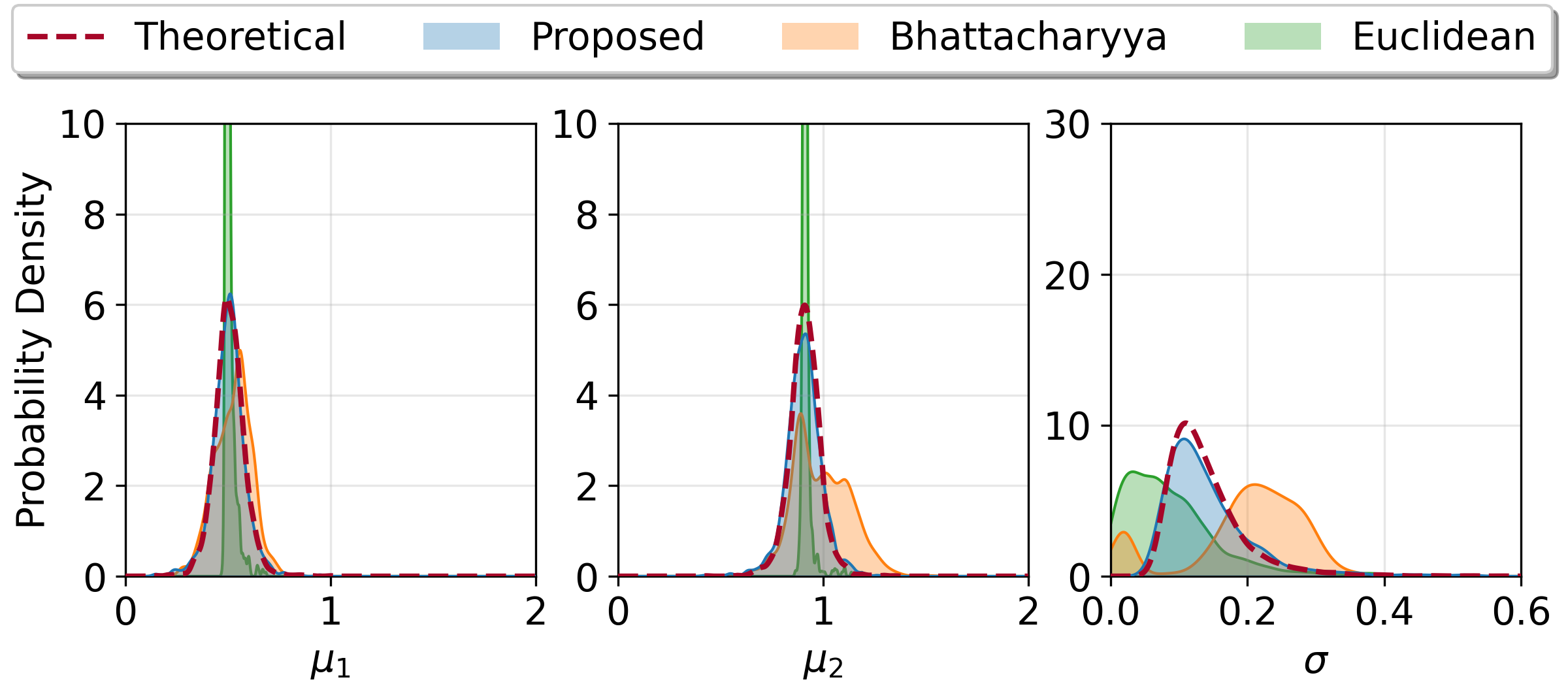}
						\subcaption{$N_{obs}=4$}
					\end{minipage}
					\begin{minipage}[t]{0.48\hsize}
						\centering
						\includegraphics[width=0.98\linewidth]{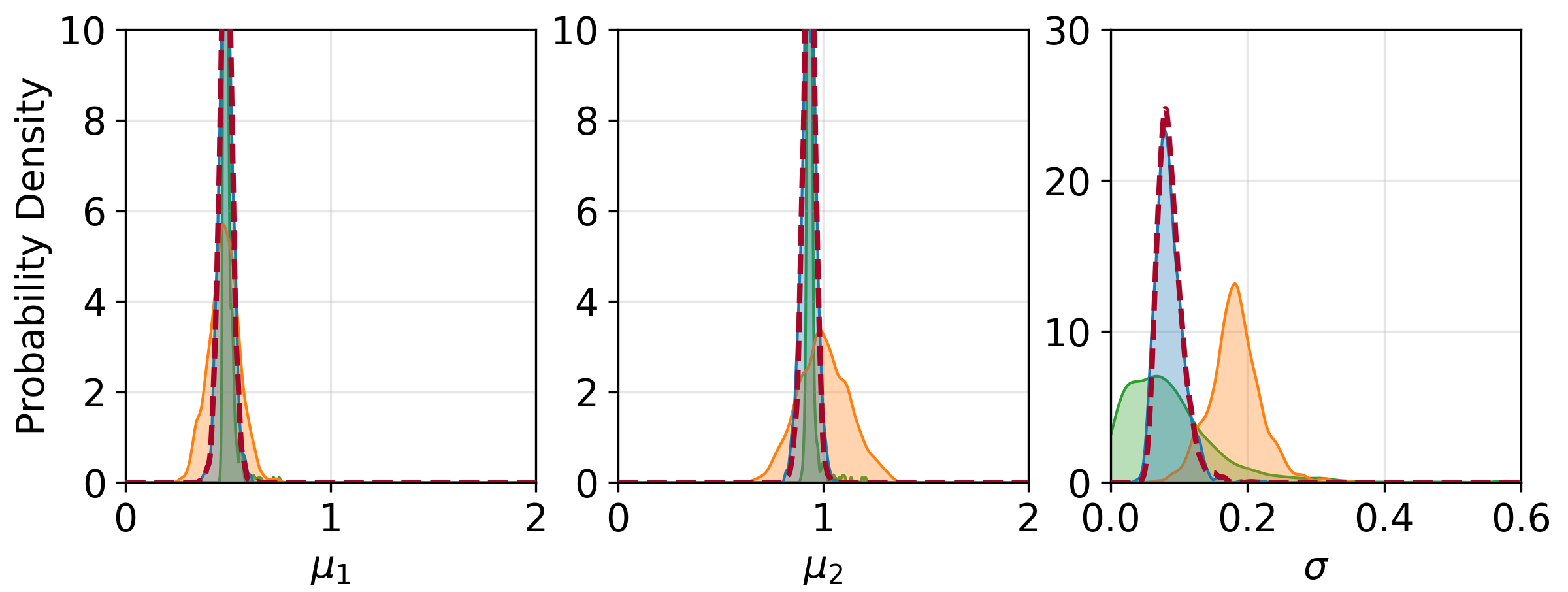}
						\subcaption{$N_{obs}=8$}
					\end{minipage} \\\\
					\begin{minipage}[t]{0.48\hsize}
						\centering
						\includegraphics[width=0.98\linewidth]{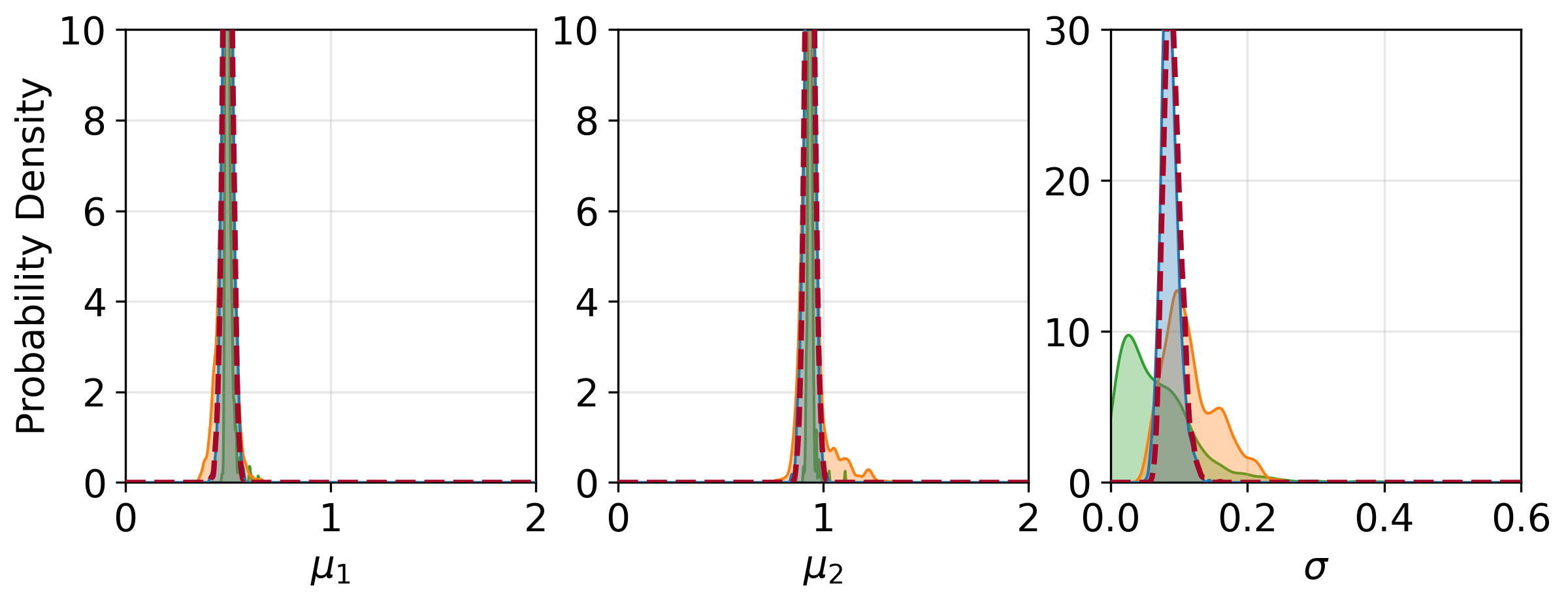}
						\subcaption{$N_{obs}=16$}
					\end{minipage}
					\begin{minipage}[t]{0.48\hsize}
						\centering
						\includegraphics[width=0.98\linewidth]{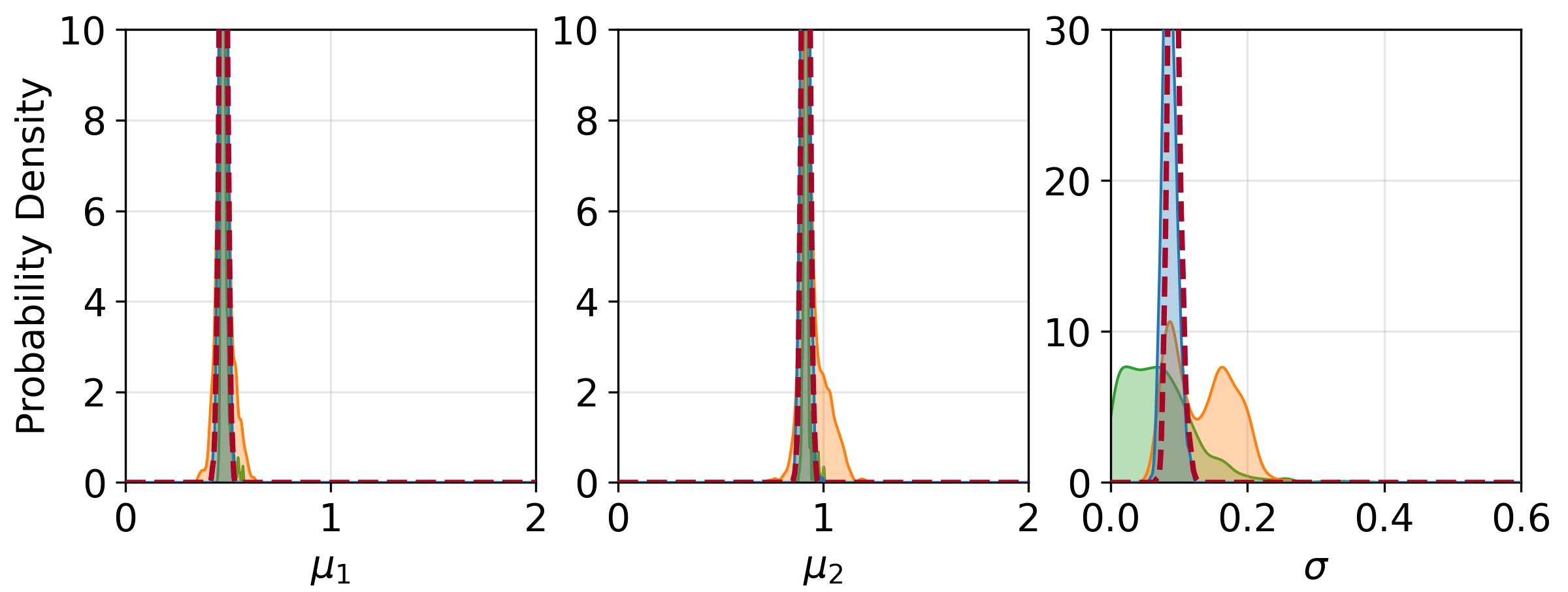}
						\subcaption{$N_{obs}=32$}
					\end{minipage}\\\\
					\begin{minipage}[t]{0.48\hsize}
						\centering
						\includegraphics[width=0.98\linewidth]{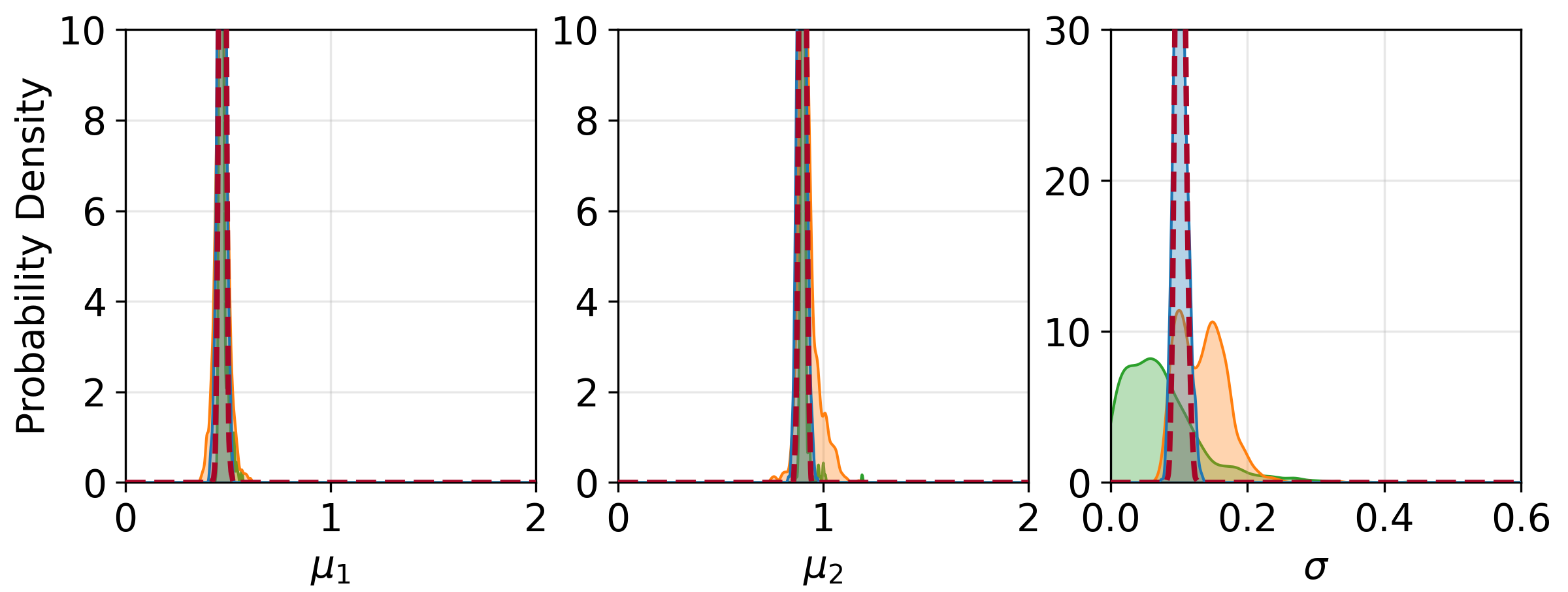}
						\subcaption{$N_{obs}=64$}
					\end{minipage}
					\begin{minipage}[t]{0.48\hsize}
						\centering
						\includegraphics[width=0.98\linewidth]{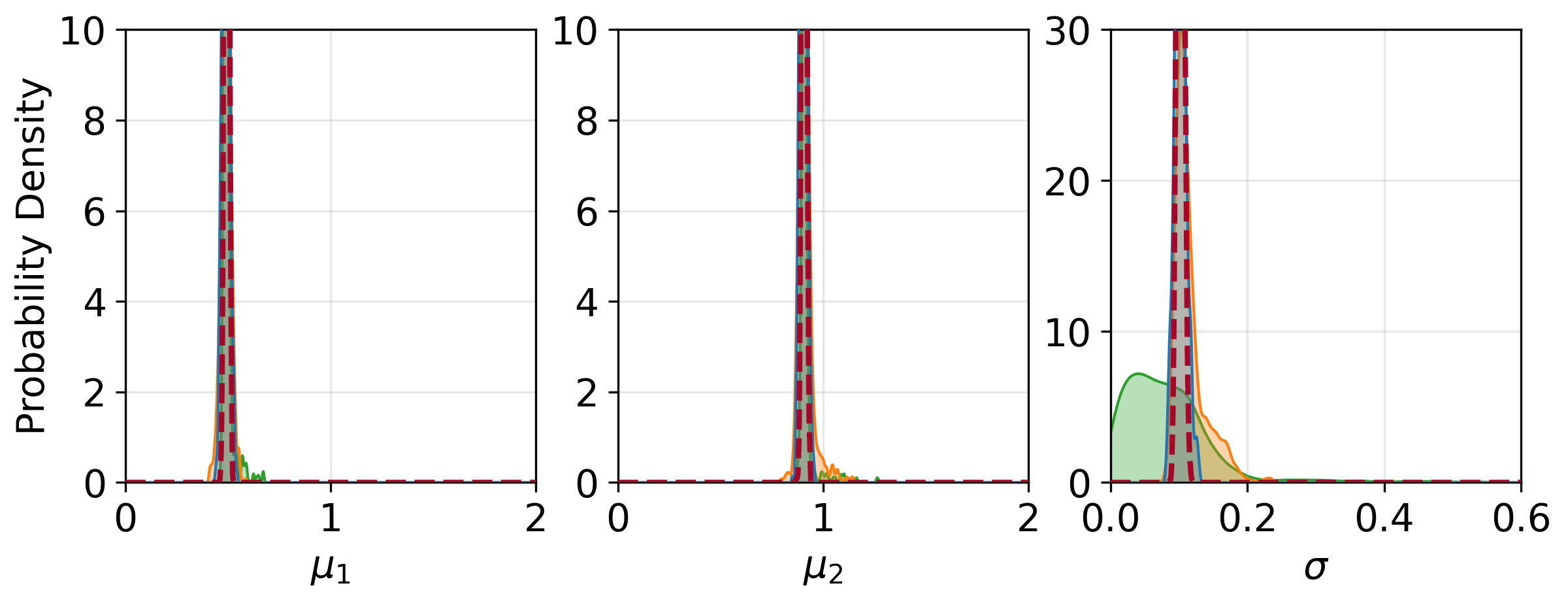}
						\subcaption{$N_{obs}=128$}
					\end{minipage}
					\\
				\end{tabular}
				\caption{Posterior distributions  of hyperparameters ($N_{sim}=128$) } \label{fig:likelihood}
			\end{figure}
			
			To further evaluate the performance of each method, a one-dimensional Gaussian kernel density estimation (KDE) was performed on the MCMC samples obtained from the theoretical likelihood and each method using \( N_{sim} = 128 \) for the latter.
			
			The KDE results provided  posterior distributions for each hyperparameter.
			The likelihood distributions for $N_{obs} = 4, 8, 16, 32, 64, 128$ are presented in Fig. \ref{fig:likelihood}.
			The distribution of the proposed method closely matches the true distribution derived from the theoretical likelihood for all hyperparameters, even when \( N_{\text{obs}} = 4 \).
			In contrast, the Bhattacharyya distance-based method showed reasonable resemblance for $\mu_1$ and $\mu_2$ but performed poorly in estimating $\sigma$ when $N_{obs}<16$.
			The Euclidean distance-based method, which minimizes the distance between sample means, failed to estimate $\sigma$ accurately, leading to smaller $\sigma$ values.
			These results confirm that the proposed method demonstrates superior likelihood estimation performance, particularly when the number of observations $N_{obs}$ is 64 or less, significantly outperforming conventional methods in these scenarios.

\begin{figure}[t]
				\begin{tabular}{cc}
					\begin{minipage}[t]{0.48\hsize}
						\centering
						\includegraphics[width=0.98\linewidth]{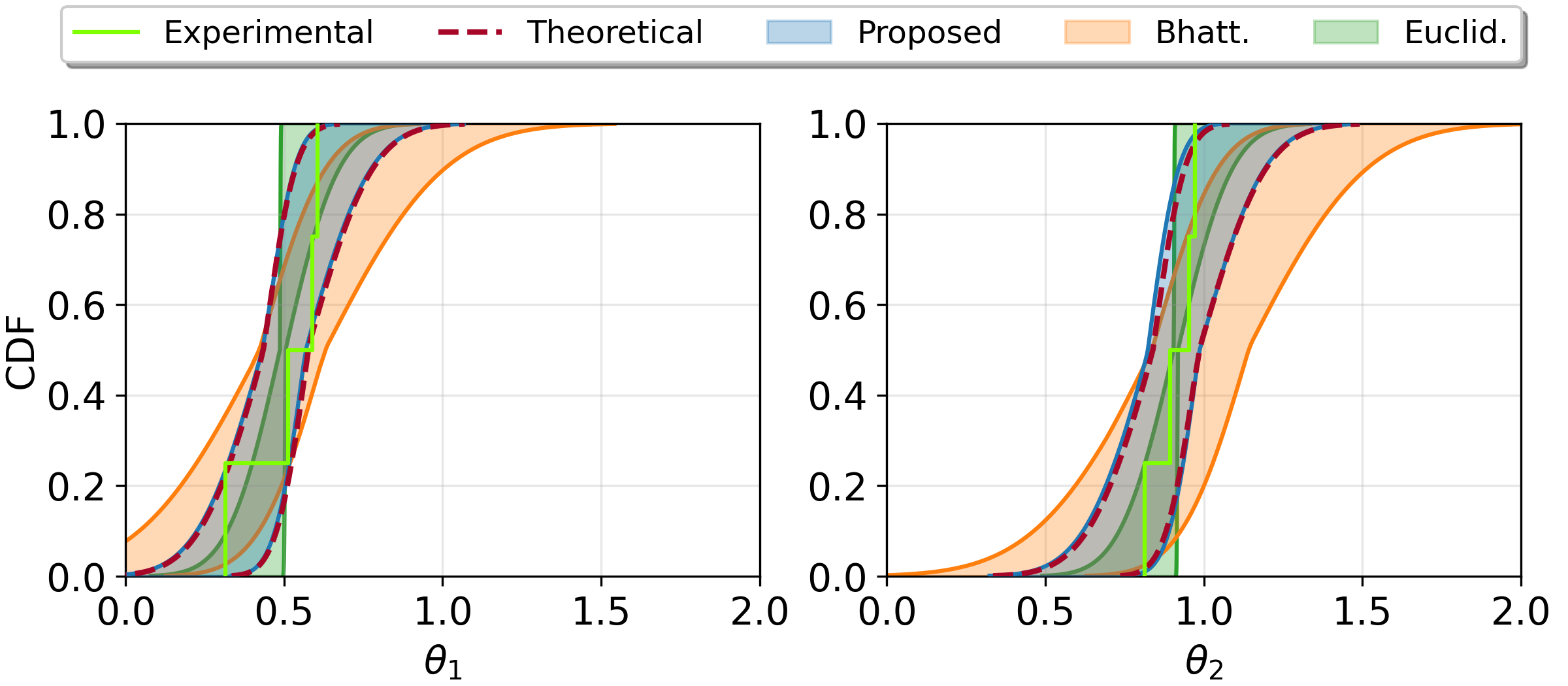}
						\subcaption{$N_{obs}=4$}
					\end{minipage}
					\begin{minipage}[t]{0.48\hsize}
						\centering
						\includegraphics[width=0.98\linewidth]{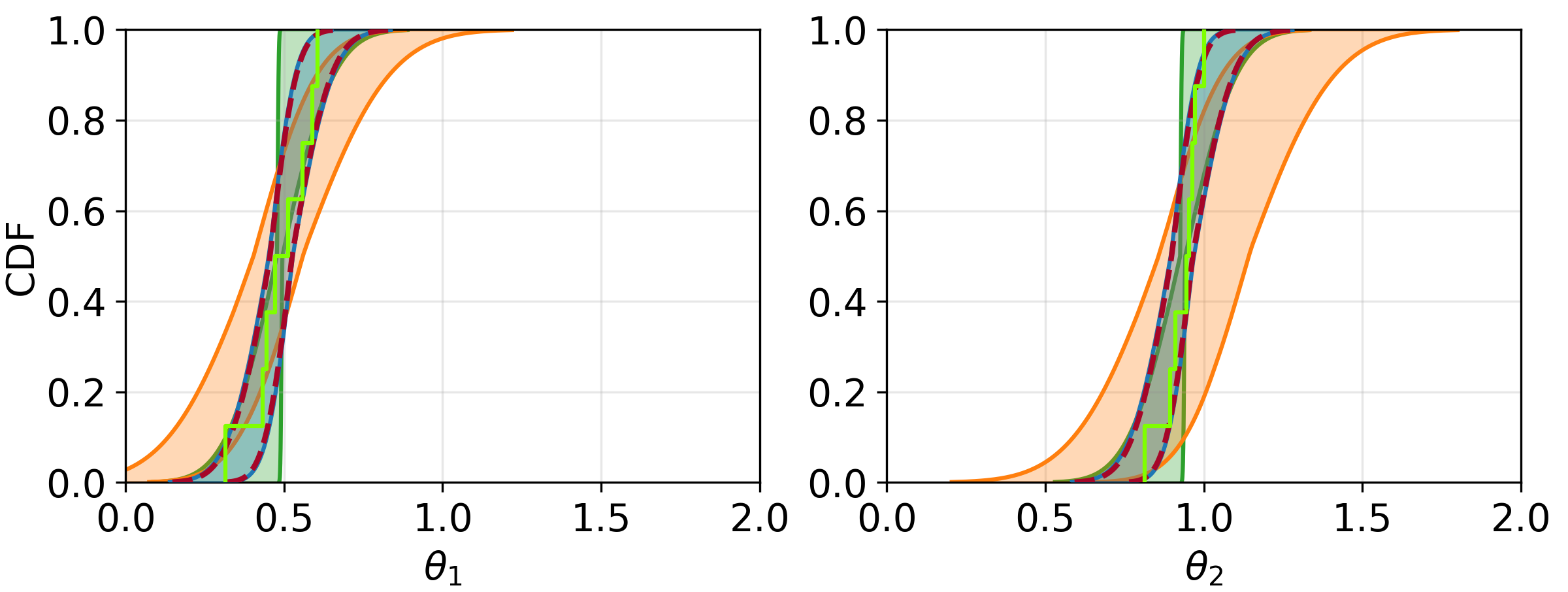}
						\subcaption{$N_{obs}=8$}
					\end{minipage} \\\\
					\begin{minipage}[t]{0.48\hsize}
						\centering
						\includegraphics[width=0.98\linewidth]{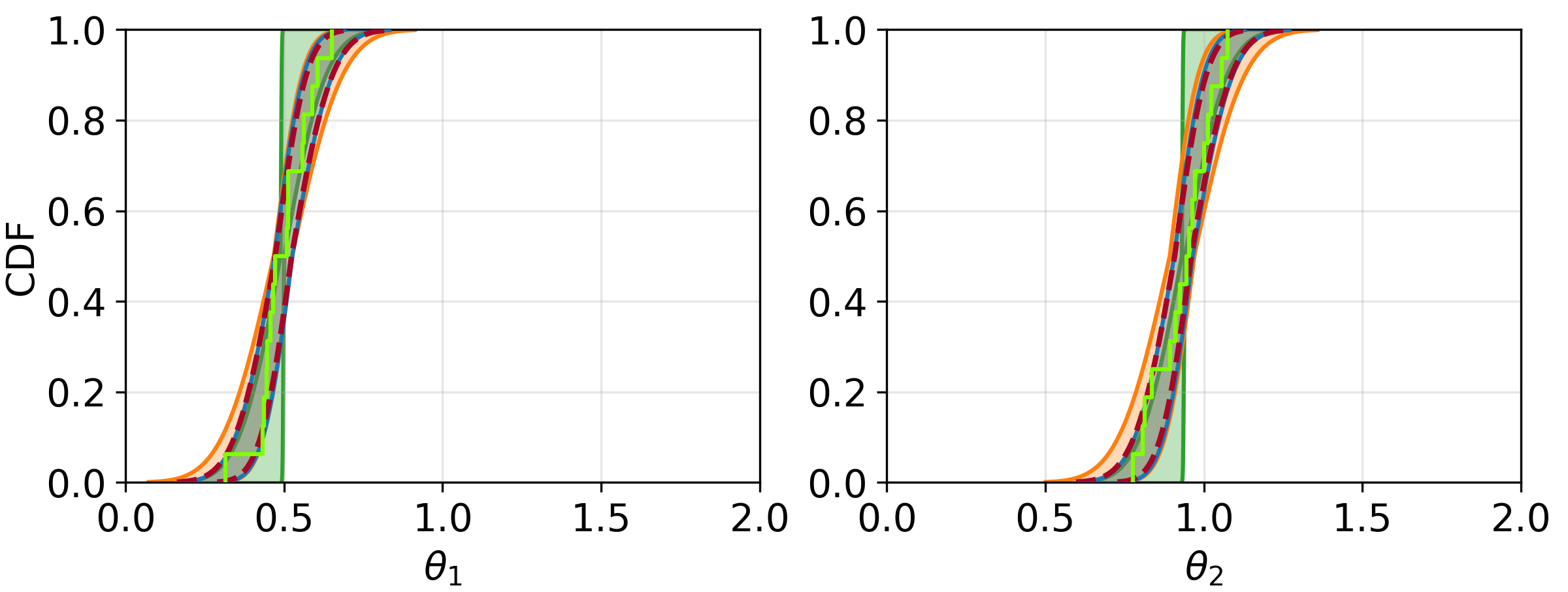}
						\subcaption{$N_{obs}=16$}
					\end{minipage}
					\begin{minipage}[t]{0.48\hsize}
						\centering
						\includegraphics[width=0.98\linewidth]{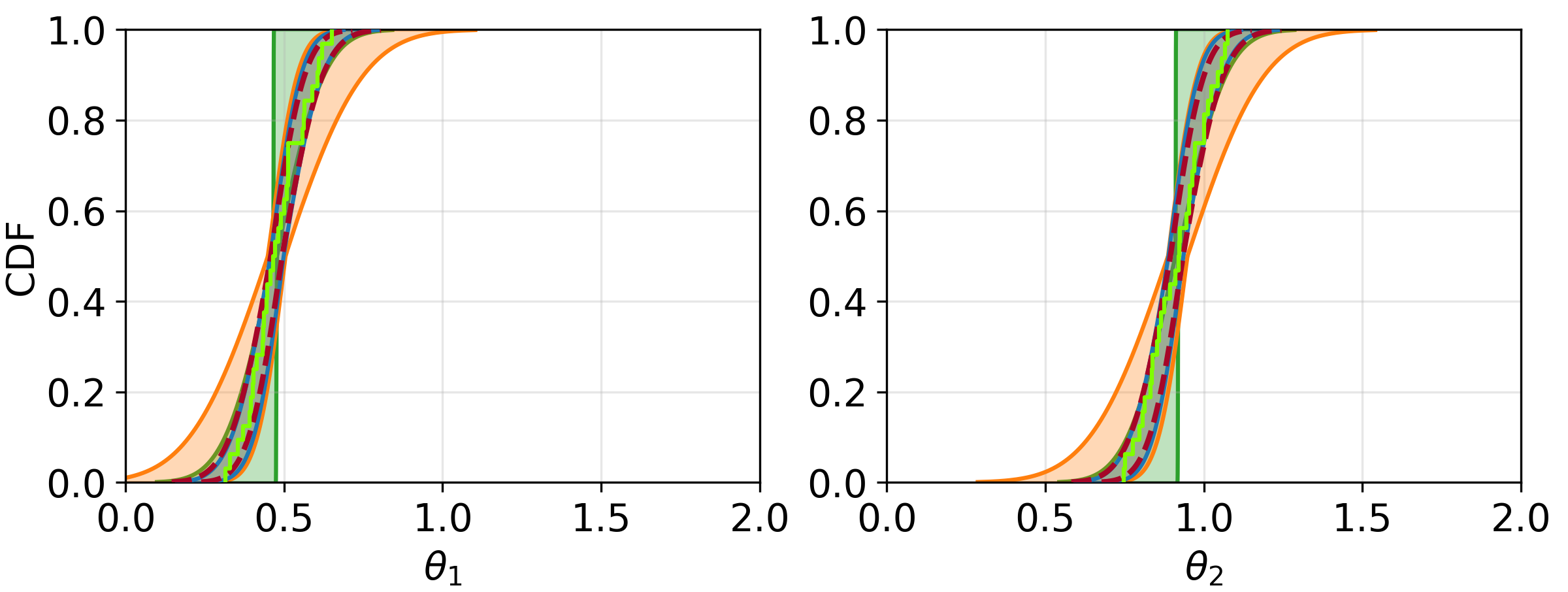}
						\subcaption{$N_{obs}=32$}
					\end{minipage}\\\\
					\begin{minipage}[t]{0.48\hsize}
						\centering
						\includegraphics[width=0.98\linewidth]{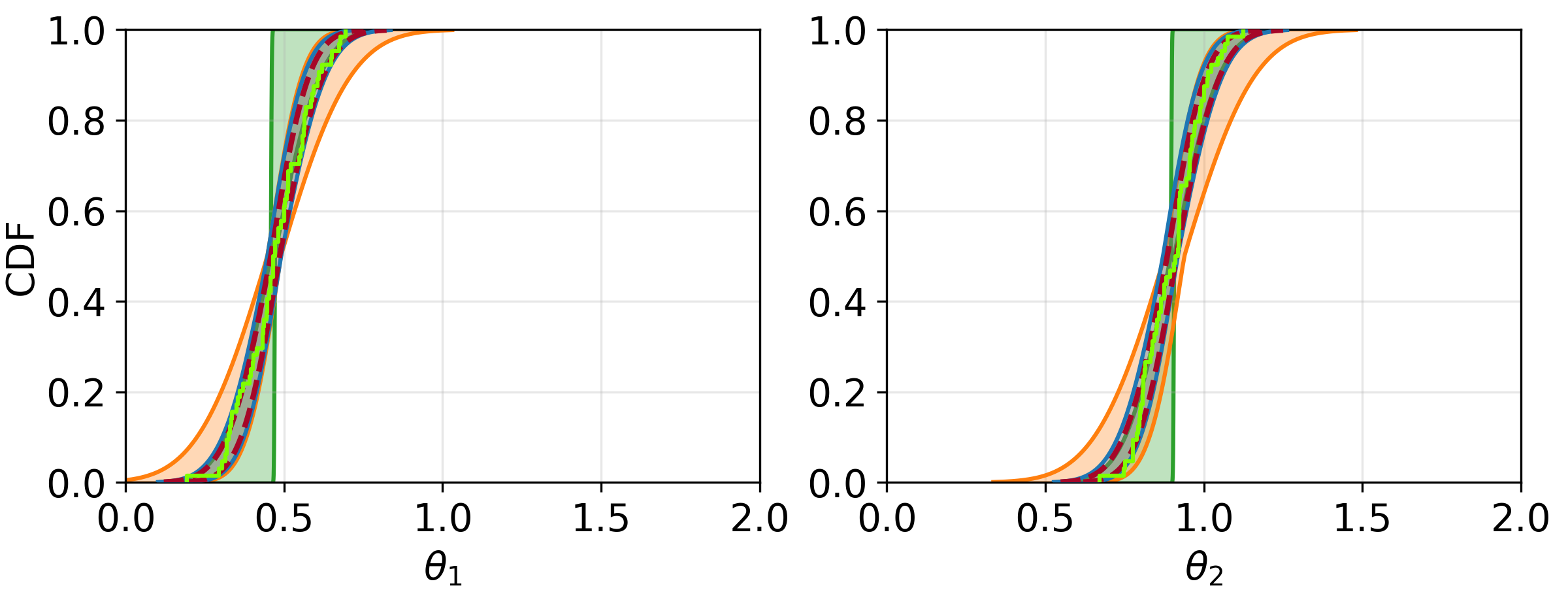}
						\subcaption{$N_{obs}=64$}
					\end{minipage}
					\begin{minipage}[t]{0.48\hsize}
						\centering
						\includegraphics[width=0.98\linewidth]{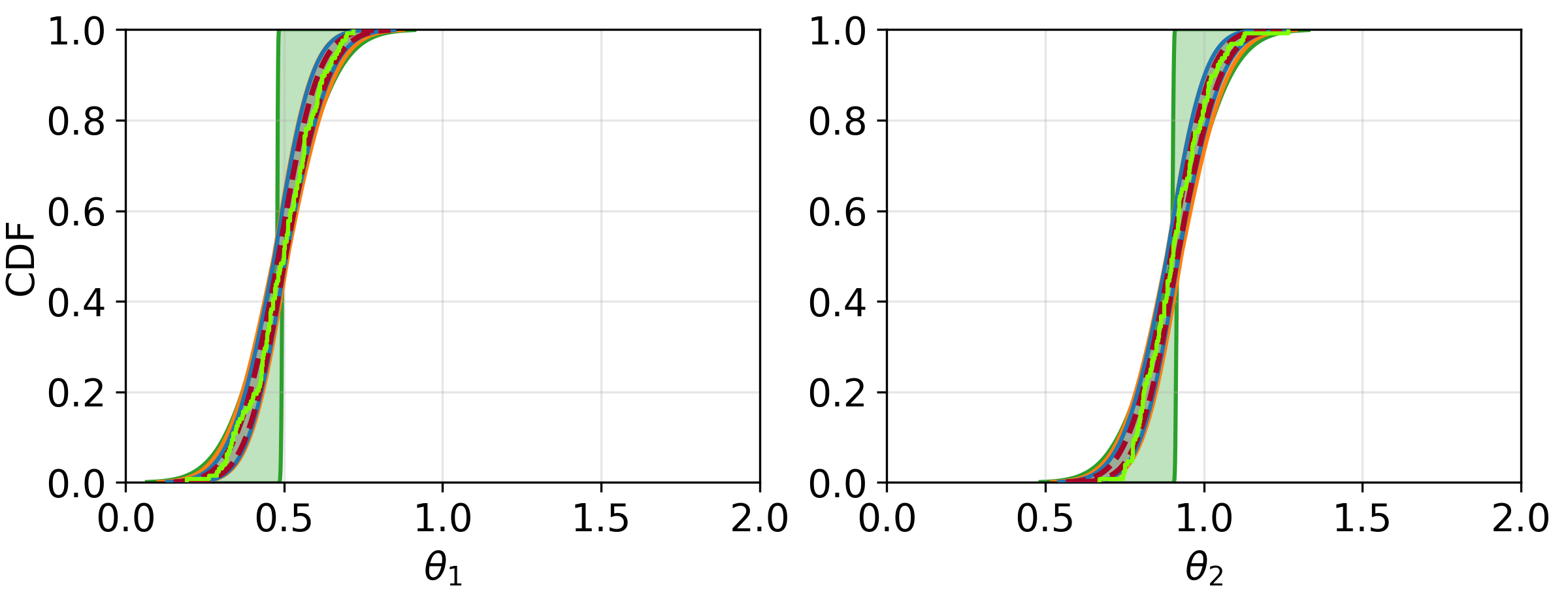}
						\subcaption{$N_{obs}=128$}
					\end{minipage} \\
				\end{tabular}
				\caption{P-boxes of the parameter $\boldsymbol{\uptheta}$ ($\alpha-\text{level} = 0.5$, $N_{sim}=128$)} \label{fig:pbox}
			\end{figure}

			Subsequently, uncertainty propagations were performed using the hyperparameters obtained from MCMC to derive the probability boxes (p-boxes) of the parameters, as shown in Fig. \ref{fig:pbox}. Each p-box corresponded to an alpha level of 0.5.
			In the figure, the term “experimental” denotes the CDF of the parameter samples employed during the generation of observation data. In practical applications, these parameter samples are not directly observable; they are included here for illustration purposes.

			All methods produced results that included the target values.
			The p-box width obtained by the proposed method was the closest to that obtained by theoretical likelihood.
			This suggests that the proposed method appropriately evaluates the uncertainty owing to the lack of information with a small \( N_{obs} \) and clearly shows a decrease in epistemic uncertainty with increasing \( N_{obs} \).
			Furthermore, with sufficient observations and simulations (\( N_{obs} = 128 \), \( N_{sim} = 128 \)), the Bhattacharyya distance-based method produced results equivalent to those of the proposed method.

		\subsection{Summary}
			In this study, the proposed latent space-based method for stochastic model updating was evaluated against traditional Bhattacharyya distance-based and Euclidean distance-based methods.
			The proposed method demonstrated superior likelihood estimation performance, particularly with limited observations (\( N_{obs} \leq 64 \)), showing lower JS divergence and higher efficiency.
			
			The proposed method effectively reduced the JS divergence with fewer simulations (\( N_{sim} \)), thereby proving to be efficient for time-consuming simulations.
			Kernel density estimation confirmed that the proposed method closely matched the theoretical likelihood distribution, outperforming traditional methods, particularly in estimating the variance parameters.
			Uncertainty propagation using MCMC-derived hyperparameters validated the effectiveness of the method, with the p-box width aligned with the theoretical likelihood results, indicating an appropriate evaluation of uncertainty with a small \( N_{obs} \) and reduced epistemic uncertainty as \( N_{obs} \) increased.
			Overall, the proposed method demonstrated superior efficiency and accuracy for stochastic model updating, making it a promising approach for applications with sparse or expensive data.

	\section{Model calibration of NASA UQ challenge 2019}
		To validate the applicability of the proposed method for high-dimensional data, the model calibration problem of the NASA UQ Challenge 2019 \cite{Crespo22, Bi22NASA, Lye21} was addressed. This experiment aims to demonstrate the effectiveness of the approach in handling complex and high-dimensional uncertainty quantification tasks.
		
		\subsection{Problem description}
			The NASA UQ challenge 2019 focused on assessing and improving the methods for quantifying uncertainty in complex systems.
			The challenge presented a series of problems that required participants to accurately estimate and manage uncertainties in the computational models.
			The goal was to advance state-of-the-art uncertainty quantification techniques.
			
			This study focused on the model calibration subproblem of the NASA UQ Challenge 2019.
			In this part of the challenge, the participants were provided with a baseline computational model and observational data.
			The task was to update the model parameters to better reflect the observed data, thereby reducing the discrepancies between model predictions and actual observations.
			The model updating problem involved estimating the true values or distributions of the parameters within the provided uncertainty bounds using the available data to improve the model's predictive accuracy.

			\begin{figure}[t]
				\centering
				\includegraphics[width=0.97\linewidth]{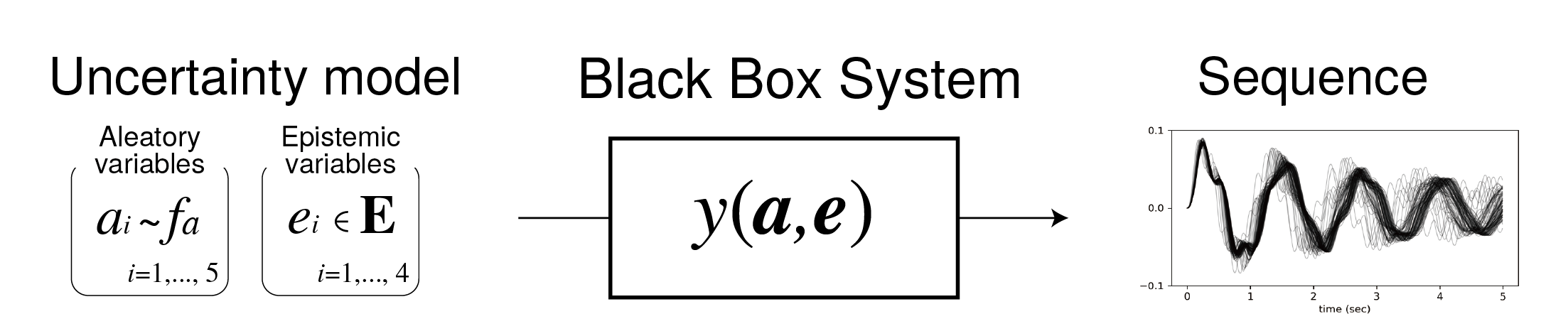}
				\caption{Subsystem of NASA challenge}\label{fig:subsystem}
			\end{figure}
			
			The subsystem provided as part of the challenge is shown in Fig. ~\ref{fig:subsystem}.
			In this figure, the subsystem is represented as \( y(\boldsymbol{a}, \boldsymbol{e}) \), where \( \boldsymbol{a} \in \mathbb{R}^{5} \) and \( \boldsymbol{e} \in \mathbb{R}^{4} \) are aleatory and epistemic variables, respectively.
			The sequence was sampled at intervals of 0.001 s over 5,001 steps.
			An epistemic variable \( \boldsymbol{e} \) is an unknown but fixed parameter, whereas  an aleatory variable \( \boldsymbol{a} \) is an unknown parameter with an unknown joint PDF \( f_a \).
			The problem of estimating the distribution of \(\boldsymbol{a}\) and the value of \( \boldsymbol{e} \) was addressed from the 100 observation sequences provided.

		\subsection{Parameterization for uncertainty model of aleatory variables}
\begin{figure}[t]
				\centering
				\includegraphics[width=0.5\linewidth]{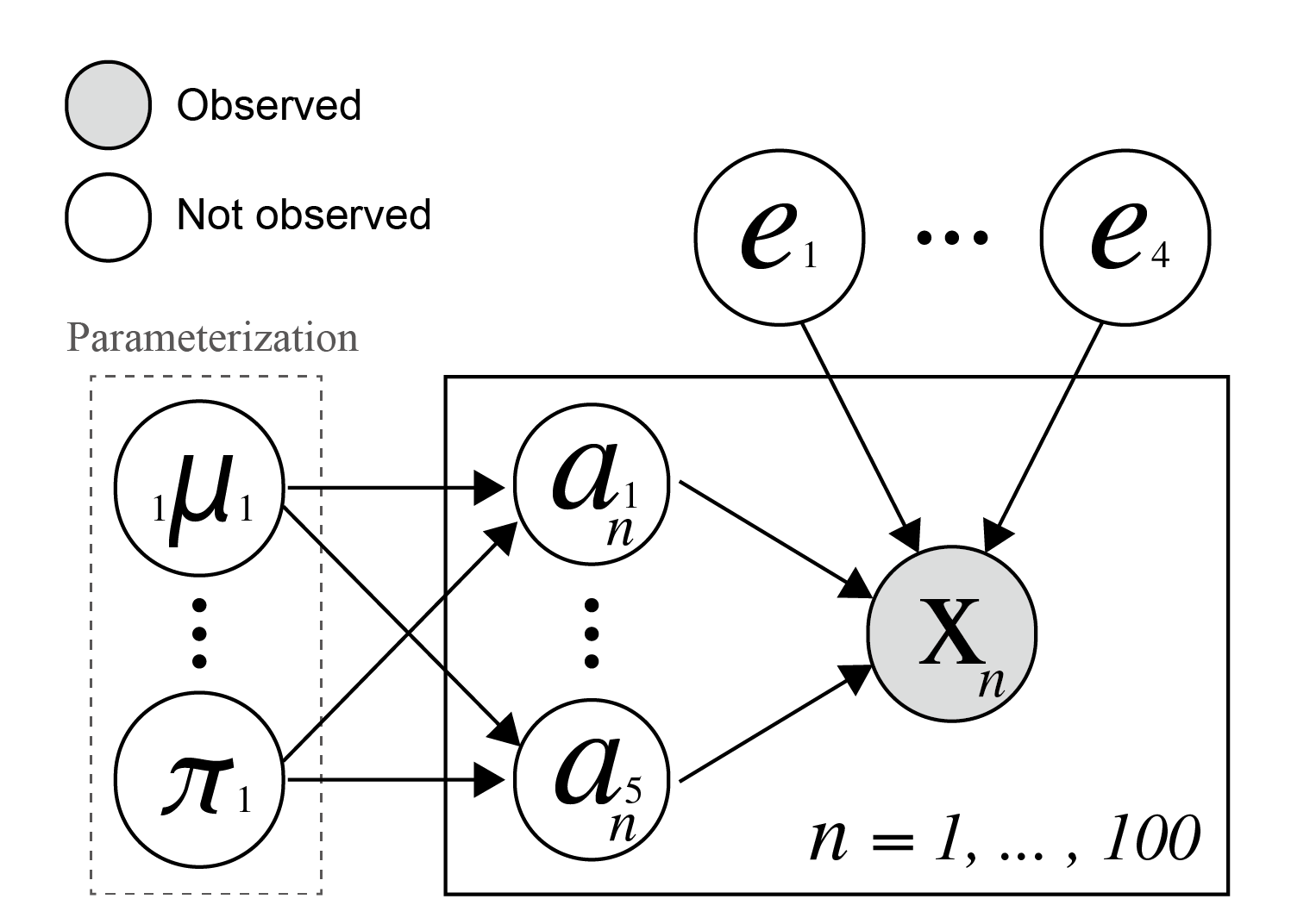}
				\caption{Graphical model representing the parameterized Gaussian mixture model}\label{fig:parameterization}
			\end{figure}

			In the absence of knowledge regarding the probability distribution \( f_a \), this study employed a Gaussian mixture model, as illustrated in Fig. \ref{fig:parameterization}.
			To model the multimodality and correlations between parameters with minimal degrees of freedom, we adopted a five-dimensional Gaussian mixture model with two components (classes \( k=1,2 \)). The probability distribution of the aleatory variables \( \boldsymbol{a} = [a_1, a_2, \dots, a_5]^T \) is expressed as follows:
			\begin{equation}
				\label{eq:GMM}
				p(\boldsymbol{a}) = \sum_{k=1}^2 \pi_k \mathcal{N}_T(\boldsymbol{a}|\boldsymbol{\mu}_k, \boldsymbol{\Sigma}_k, A)
			\end{equation}
			\noindent where \(\pi_k\) is the mixing coefficient of the \(k\)th Gaussian component, and the sum of \(\pi_k\) over all components is 1, ensuring a valid probability distribution. \(\mathcal{N}_T(\boldsymbol{a}|\boldsymbol{\mu}_k, \boldsymbol{\Sigma}_k, A)\) represents a truncated Gaussian distribution, where \( \boldsymbol{a} \) is restricted to the truncation range \( A \), defined as \( A = \{ \boldsymbol{a} \in \mathbb{R}^5 \mid 0 \leq a_i \leq 2, \, i = 1, 2, \dots, 5 \} \). This truncation ensured that the Gaussian mixture model is defined only within the bounds provided by the NASA challenge.
			In this model, \( \boldsymbol{\mu}_k \) and \( \boldsymbol{\Sigma}_k \) represent the mean vector and covariance matrix of the \( k \)th Gaussian component, respectively, and are given by
			\begin{equation}
				\label{eq:GMM_parm}
				\boldsymbol{\mu}_k = \begin{bmatrix}
				{}_{1}\mu_{k} \\
										 {}_{2}\mu_{k} \\
										 \vdots \\
										 {}_{5}\mu_{k}
				\end{bmatrix},
				\quad
				\boldsymbol{\Sigma}_k = \begin{bmatrix}
				{}_{1}\sigma_{k}^2 & {}_{12}\rho_{k} \cdot {}_{1}\sigma_{k} \cdot {}_{2}\sigma_{k}     & \cdots & {}_{15}\rho_{k} \cdot {}_{1}\sigma_{k}\cdot {}_{5}\sigma_{k} \\
											{}_{12}\rho_{k} \cdot {}_{1}\sigma_{k}\cdot {}_{2}\sigma_{k}  & {}_{2}\sigma_{k}^2 & \cdots & {}_{25}\rho_{k}  \cdot {}_{2}\sigma_{k}\cdot {}_{5}\sigma_{k}  \\
											\vdots           & \vdots             & \ddots & \vdots \\
											{}_{15}\rho_{k}\cdot {}_{1}\sigma {}_{k}\cdot {}_{5}\sigma_{k}  & {}_{25}\rho_{k} \cdot {}_{2}\sigma_{k} \cdot {}_{2}\sigma_{k}    & \cdots & {}_{5}\sigma_{k}^2
				\end{bmatrix}
			\end{equation}
			
			The parameterized model included 41 hyperparameters: 10 mean values \({}_{1}\mu_{1}, \dots, {}_{5}\mu_{2}\), 10 variance terms \({}_{1}\sigma_{1},\dots, {}_{5}\sigma_{2}\), 20 correlation coefficients \({}_{12}\rho_{1}, \dots, {}_{45}\rho_{2}\), and one mixing coefficient \(\pi_1\). Moreover, the problem involved estimating four epistemic variables \(e_1, \dots, e_4\), resulting in 45 parameters and hyperparameters.

		\subsection{Training of VAE}
			The dataset used for training the VAE was created by generating uniform random numbers within the parameter bounds \([0, 2]\), provided as prior information in the challenge.
			To enhance computational efficiency, the sequence was downsampled to 0.01 s over 501 steps, with 11 zero steps added at the beginning of the sequence. This resulted in the alignment of the data in 512 dimensions.
			The dataset was structured as a tensor with dimensions of (100,000, 1, 1, 512). Using this generated dataset, a VAE with the same structure as in the previous section but with six latent dimension was trained.

			\begin{table}[b]
			    \caption{Intervals of Parameters}\label{tab:interval}
			    \centering
			    \begin{tabular}{cc}
			        \hline
			          Parameter  &  interval    \\
			        \hline
					${}_i\sigma_k^2$ &    $(0, \infty)$            \\
					${}_{ij}\rho_k$ &    $[-1, -1]$            \\
					$\pi_1$ &    $[0.5, 1]$            \\
			        \hline
			    \end{tabular}
			\end{table}

		\subsection{Sampling with MCMC}
			Following the approach described in the previous section, MCMC sampling was conducted using the replica exchange Monte Carlo method with non-informative prior distributions.
			In this experiment, the number of simulations  $N_{sim}$ was set to 1,000. Sampling was conducted within the intervals listed in Table \ref{tab:interval}.
			
			Simultaneously updating all 45 parameters is challenging.
			Therefore, the strategy involved initially updating 25 parameters: the means \({}_{1}\mu_{1}, \dots, {}_{5}\mu_{2}\), standard deviations \({}_{1}\sigma_{1},\dots, {}_{5}\sigma_{2}\), mixing coefficient \(\pi_1\), and epistemic variables \(e_1, \dots, e_4\).
			After 1,000 iterations (with 120 samples per iteration) of MCMC for these 25 parameters, all 45 parameters, including the 20 correlation coefficients \({}_{12}\rho_{1}, \dots, {}_{45}\rho_{2}\), were updated simultaneously.
			The initial values for the 20 correlation terms were set to 0, and an additional 1,000 iterations (with 180 samples per iteration) were performed.
			All samples, except those from the last 700 iterations, were discarded as burn-ins. Thinning was performed by selecting every 45th sample, resulting in  2,800 final samples.

		\subsection{Results of stochastic model updating}
\begin{figure}[H]
				\centering
				\includegraphics[width=0.83\linewidth]{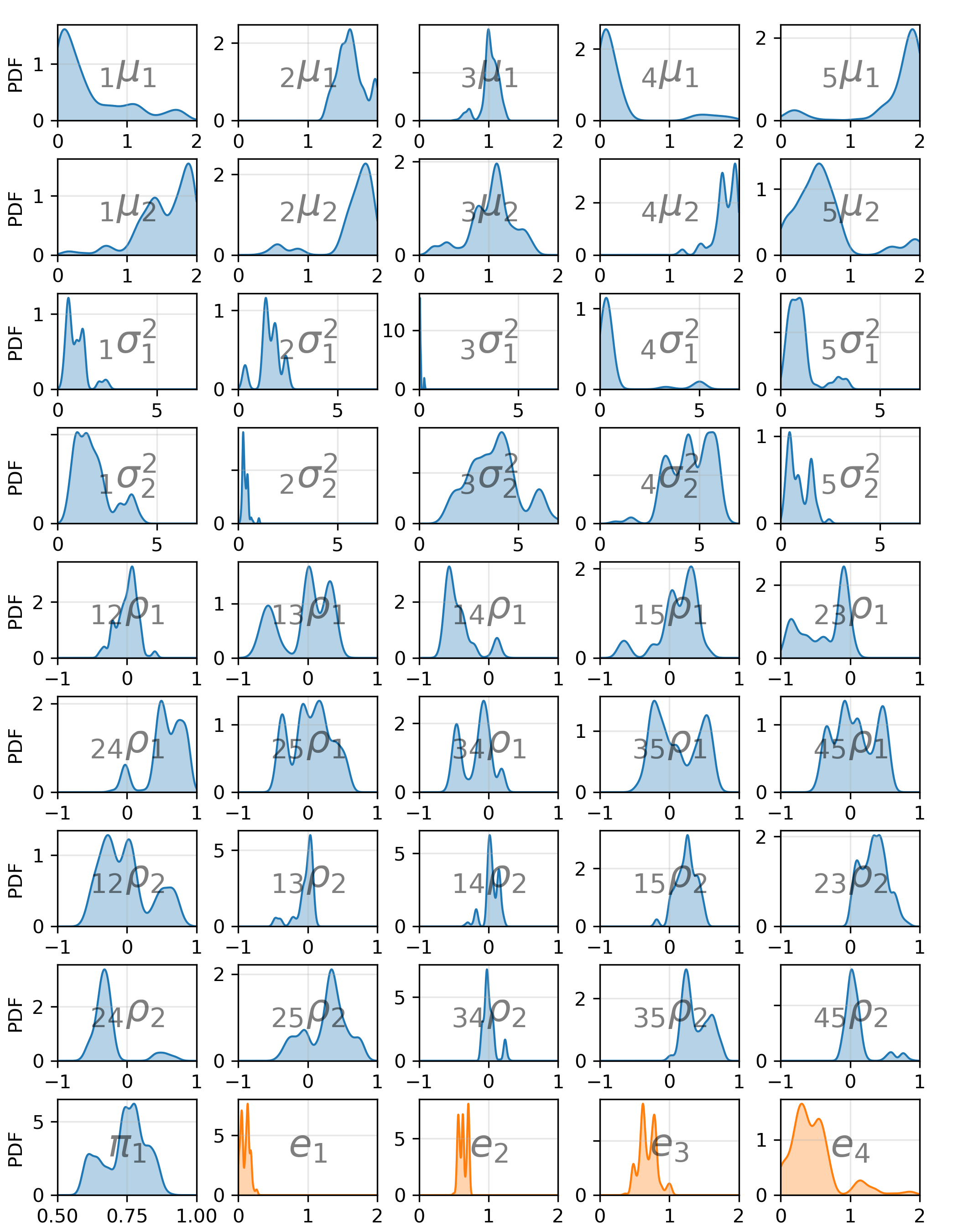}
				\caption{Posterior distributions of parameters}\label{fig:postdist}
			\end{figure}

			As a result of the Bayesian updating process, the posterior distributions of the 41 hyperparameters and four parameters were estimated using kernel density estimation, as shown in Fig. \ref{fig:postdist}.\\
			
\begin{figure}[t]
				\centering
				\includegraphics[width=0.75\linewidth]{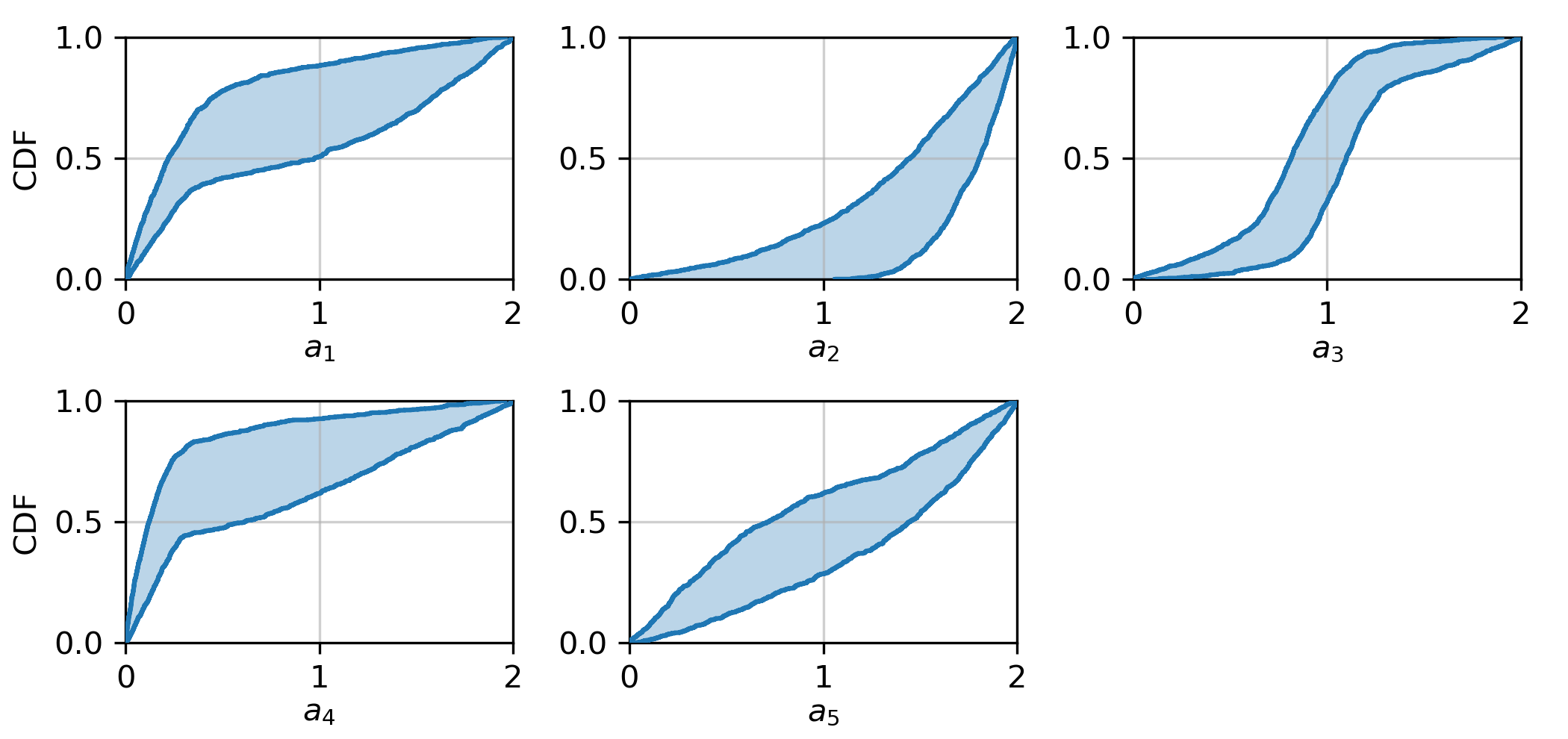}
				\caption{P-boxes for aleatory variables ($\alpha-\text{level} = 0.9$)}\label{fig:pbox_aleatory}
			\end{figure}

			The p-boxes of the aleatory variables \(\boldsymbol{a}\) shown in Fig. \ref{fig:pbox_aleatory}, demonstrate a significant narrowing compared to the initial bounds, indicating a reduction in uncertainty through the  Bayesian model updating process.

\begin{figure}[t]
				\centering
				\includegraphics[width=0.8\linewidth]{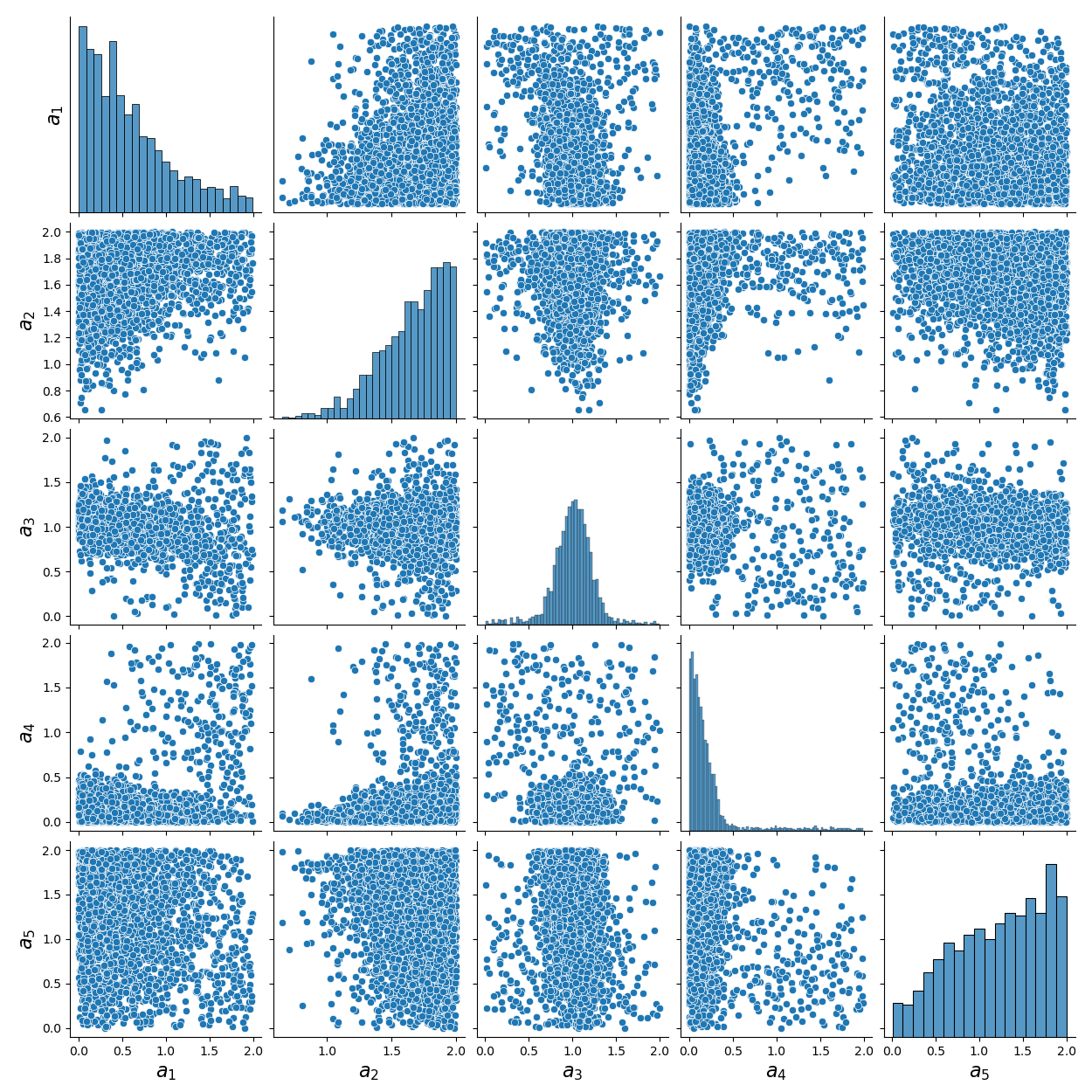}
				\caption{Three thousand samples of the aleatory variables from the  maximum a posteriori estimation}\label{fig:pos_ap}
			\end{figure}

			The maximum a posteriori (MAP) estimates of the aleatory hyperparameters, where the likelihood reaches its maximum in Bayesian updating, are listed in Table \ref{tab:mlvalue}.
			Samples of the estimated distribution of the aleatory variables based on the  MAP estimation of the 41 hyperparameters in Table \ref{tab:mlvalue} are shown in Fig. \ref{fig:pos_ap}.

			\begin{table}[t]
				\caption{Maximum a posteriori estimates of aleatory hyperparameters}\label{tab:mlvalue}
				\centering
				\begin{tabular}{cc|cc|cc|cc|cc}
					\hline
					\multicolumn{2}{c|}{$\boldsymbol{\mu}_1$} & \multicolumn{2}{c|}{$\boldsymbol{\mu}_2$} & \multicolumn{2}{c|}{$\boldsymbol{\sigma}_1^2$} & \multicolumn{2}{c|}{$\boldsymbol{\sigma}_2^2$} & \multicolumn{2}{c}{$\pi_1$} \\
					\hline
					${}_{1}\mu_{1}$   & 0.0507  & ${}_{1}\mu_{2}$   & 1.8747  & ${}_{1}\sigma_{1}^2$ & 0.9904  & ${}_{1}\sigma_{2}^2$ & 0.7396 & $\pi_1$ & 0.8592 \\
					${}_{2}\mu_{1}$   & 1.6690  & ${}_{2}\mu_{2}$   & 1.9012  & ${}_{2}\sigma_{1}^2$ & 1.2857  & ${}_{2}\sigma_{2}^2$ & 0.2372 & & \\
					${}_{3}\mu_{1}$   & 1.0744  & ${}_{3}\mu_{2}$   & 1.1588  & ${}_{3}\sigma_{1}^2$ & 0.0444  & ${}_{3}\sigma_{2}^2$ & 3.9733 & & \\
					${}_{4}\mu_{1}$   & 0.1672  & ${}_{4}\mu_{2}$   & 1.9494  & ${}_{4}\sigma_{1}^2$ & 0.2558  & ${}_{4}\sigma_{2}^2$ & 4.1211 & & \\
					${}_{5}\mu_{1}$   & 1.4526  & ${}_{5}\mu_{2}$   & 0.7430  & ${}_{5}\sigma_{1}^2$ & 1.1601  & ${}_{5}\sigma_{2}^2$ & 0.4426 & & \\
					\hline
					\hline
					\multicolumn{10}{c}{$\boldsymbol{\rho}_1$} \\
					\hline
					${}_{12}\rho_{1}$ & -0.2284 & ${}_{13}\rho_{1}$ & -0.5190 & ${}_{14}\rho_{1}$ & -0.5611 & ${}_{15}\rho_{1}$ & 0.0273 & ${}_{23}\rho_{1}$ & -0.1119 \\
					${}_{24}\rho_{1}$ & 0.8756  & ${}_{25}\rho_{1}$ & -0.4133 & ${}_{34}\rho_{1}$ &  0.0000 & ${}_{35}\rho_{1}$ & -0.2521 & ${}_{45}\rho_{1}$ & -0.1698 \\
					\hline
					\hline
					\multicolumn{10}{c}{$\boldsymbol{\rho}_2$} \\
					\hline
					${}_{12}\rho_{2}$ & -0.2568 & ${}_{13}\rho_{2}$ &  0.0268 & ${}_{14}\rho_{2}$ &  0.1516 & ${}_{15}\rho_{2}$ &  0.1383 & ${}_{23}\rho_{2}$ &  0.5960 \\
					${}_{24}\rho_{2}$ & -0.2449 & ${}_{25}\rho_{2}$ &  0.3076 & ${}_{34}\rho_{2}$ & -0.0808 & ${}_{35}\rho_{2}$ &  0.1486 & ${}_{45}\rho_{2}$ & -0.0936 \\
					\hline
				\end{tabular}
			\end{table}

			\begin{table}[t]
				\caption{Maximum a posteriori estimates and reduced ranges of epistemic variables}\label{tab:mlvalue_e}
				\centering
				\begin{tabular}{c|c|c}
					\hline
					Parameter &  Maximum a posteriori estimate & Reduced range  \\
					\hline
					\(e_1\)        & 0.120              & [0.0, 0.214]   \\
					\(e_2\)        & 0.717              & [0.526, 0.739] \\
					\(e_3\)        & 0.834              & [0.436, 1.024] \\
					\(e_4\)        & 0.683              & [0.0, 1.267]   \\
					\hline
				\end{tabular}
			\end{table}
			
			\begin{figure}[t]
				\centering
				\includegraphics[width=0.8\linewidth]{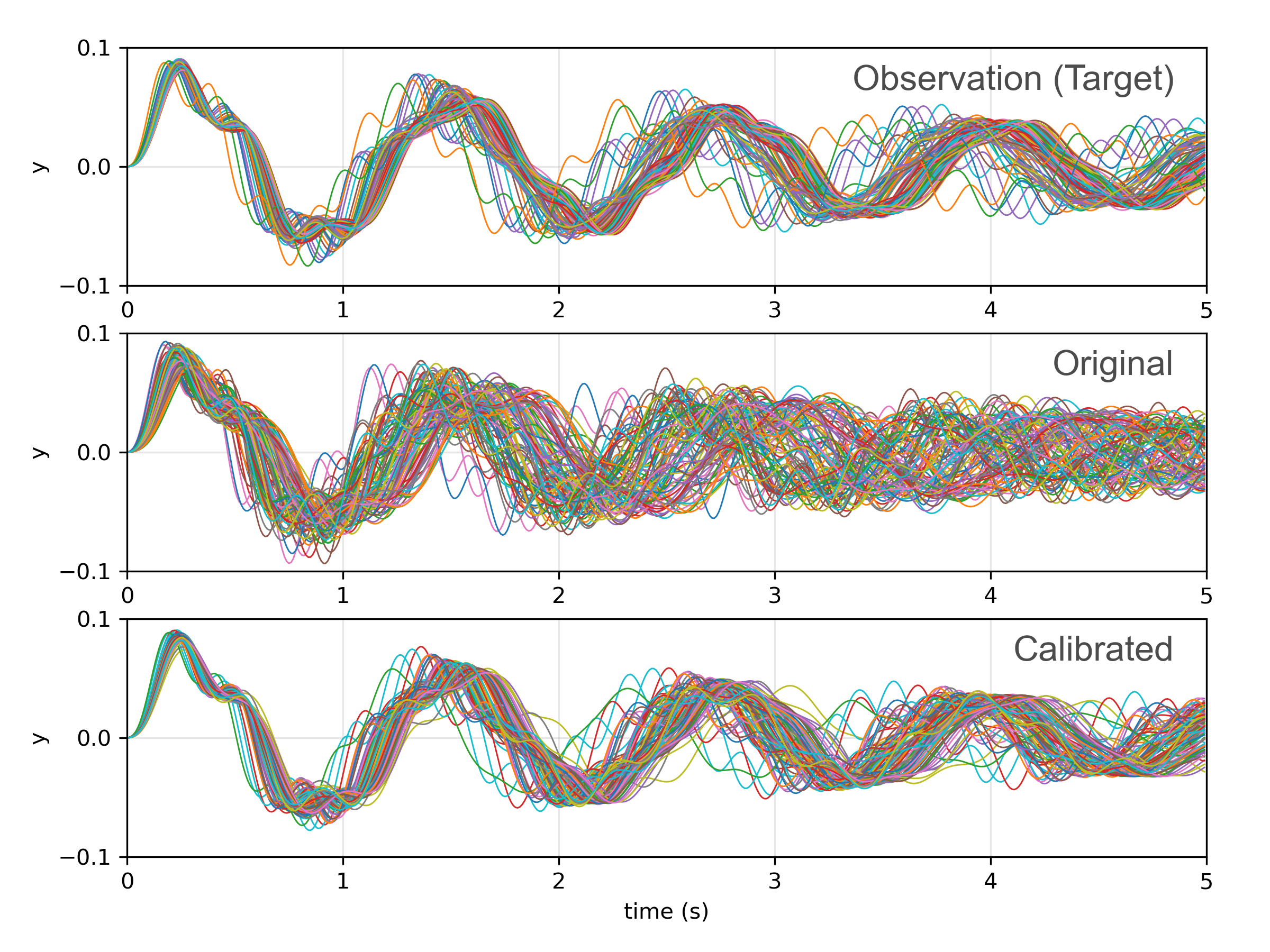}
				\caption{Comparison between the observations, original sequences, and calibrated sequences}\label{fig:comp_seq}
			\end{figure}

			\begin{figure}[t]
				\centering
				\includegraphics[width=0.8\linewidth]{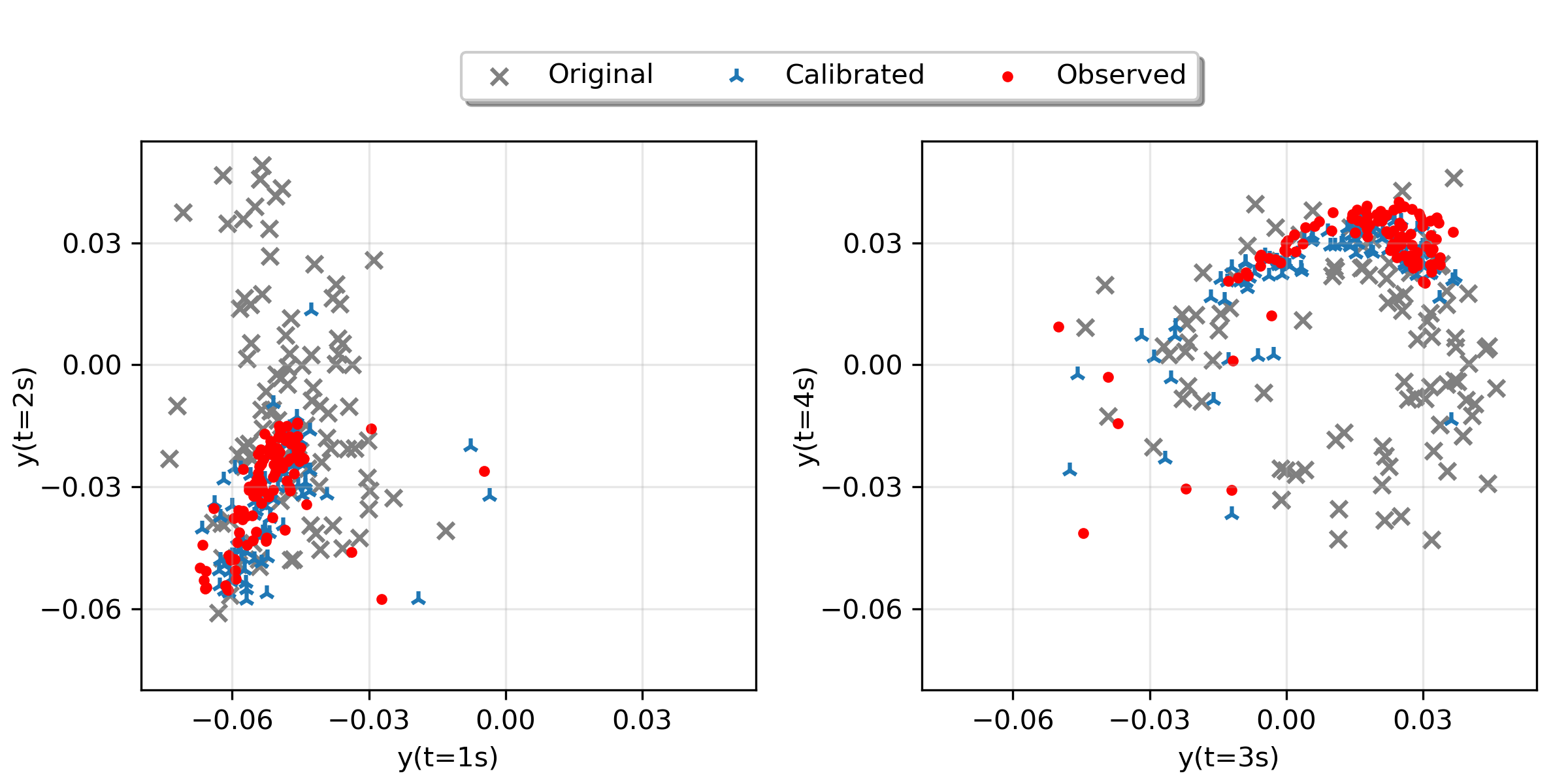}
				\caption{Scatters of the original, calibrated, and observation values of the sequences at specific time points.}\label{fig:scatter}
			\end{figure}
			
			The  MAP estimates, along with the reduced ranges of the epistemic variables, where the normalized PDF reaches 0.1, are presented in Table \ref{tab:mlvalue_e}.
			By examining the reduced ranges of the epistemic variables, it is evident that the ranges for all parameters narrowed compared to the prior range of [0, 2].
			This indicates a significant update in the posterior distribution caused by the data. Notably, the sensitivity of the model to the sequences \(e_1\) and \(e_2\) was high, as evidenced by the relatively narrow reduced ranges.
			Contrarily, \(e_4\) exhibited low sensitivity to the sequence, resulting in a much broader retained range, reflecting greater uncertainty or less influence of the data on this parameter.

			Fig. \ref{fig:comp_seq} shows the sequences of the calibrated model by MAP estimation, along with the original sequences derived from the parameters within the provided bounds and the target observations for comparison.
			The proposed method successfully generated sequences that closely matched the observed data in form and distribution even when a simple Gaussian mixture model was assumed for \( \boldsymbol{a} \).
			In addition, \ref{sec:app2} presents 20 sample sequences obtained using the same hyperparameters as those in  Fig. \ref{fig:comp_seq}.

			The scatterplots in Fig. \ref{fig:scatter} illustrate the distribution of the original, calibrated, and observed values of the sequences at specific time points.
			The left panels display two-dimensional scatter plots at 1 and 2 s, whereas the right panels show scatter plots at 3 and 4 s.
			The original values, represented by gray crosses, exhibited a wide distribution, indicating the initial uncertainty in the model.
			By contrast, the calibrated values (black triangles) were closely aligned with the observed values (red circles), demonstrating a significant reduction in the distribution spread. This indicates that the model calibration effectively reduced the discrepancies between the model predictions and observed data.
			Furthermore, the calibration results matched the main peak of the distribution and accurately captured its tails. This comprehensive alignment suggests that the proposed method successfully handled both the central and peripheral data points.

		\subsection{Summary}
			The NASA UQ Challenge 2019 presented a complex model calibration task that focused on quantifying and managing the uncertainties in computational models. This study addresses subproblem A of the challenge, where the goal is to update the model parameters using observational data to reduce the discrepancies between model predictions and actual observations.
			The proposed method involves parameterizing the probability distribution of aleatory variables using a Gaussian mixture model, resulting in a comprehensive model with 45 parameters and hyperparameters. A VAE with Bayesian updating was employed to estimate the posterior distributions of these parameters, significantly reducing the initial uncertainty of the model.
			The results demonstrated that the calibrated model closely aligned with the observed data, matching the main peak and accurately capturing the tails of the distribution. This indicates that the method effectively handles high-dimensional data and complex uncertainty quantification tasks, improving the predictive accuracy of the model.
			Overall, this experiment validates the applicability and effectiveness of the proposed method for addressing high-dimensional model calibration problems, demonstrating its potential for broader applications in uncertainty quantification.

	\section{Conclusion}
		This study presented a novel latent space-based method for stochastic model updating with limited observations and simulations.
		By leveraging a VAE, the proposed method effectively quantified uncertainties in high-dimensional data with fewer data requirements.
		Through numerical experiments on a two-degree-of-freedom shear spring model, the proposed method showed superior efficiency and accuracy compared to traditional metrics, such as Bhattacharyya and Euclidean distances.
		In addition, the applicability of this method to solving the NASA UQ challenge 2019 demonstrated the feasibility of using this method for handling high-dimensional data.
		
		The results underscore the potential of the latent space-based method in practical engineering applications, providing a robust framework for uncertainty quantification with fewer data requirements, and demonstrating its effectiveness in handling high-dimensional data.
		The ability of this method to maintain accuracy with a smaller number of observations and simulations is particularly advantageous for complex engineering systems where data collection and simulations can be time-consuming and costly.
		
		Future studies will explore enhancements to further improve the proposed method.
		One area of development is to reduce the need to perform \(N_{\text{sim}}\) simulations by integrating the surrogate models with a multimodal VAE, which can accelerate the simulation process, as suggested in \cite{Itoi24}.
		Additionally, while generating a dataset for training the VAE involves substantial computational costs, implementing sequential adaptive training methods can make this process more efficient and cost-effective.

		\appendix

	\section{Analytical computation of integral components in likelihood estimation}\label{sec:app0}
		In this Appendix, we provide a detailed explanation of the analytical computation of an integral part of Eq. (\ref{eq:smu_lik}) presented in the main text as follows:
		\begin{equation}
			L(\mathbf{X}_{\text{obs}}|\boldsymbol{\upphi}) =  \prod_{i=1}^{N_{obs}} \sum_{j=1}^{N_{sim}} \int_{\mathcal{Z}} \frac{q(\mathbf{z}| \boldsymbol{\uptheta}_j^{*}) \ q(\mathbf{z}| \mathbf{x}_{\text{obs}}^{(i)})}{q(\mathbf{z})} \ d\mathbf{z}
		\end{equation}
		First, the integral is decomposed into the individual dimensions of the latent variable \(\mathbf{z}\).
		\begin{equation}
			L(\mathbf{X}_{\text{obs}}|\boldsymbol{\upphi}) = \ \prod_{i=1}^{N_{obs}} \sum_{j=1}^{N_{sim}} \prod_{k=1}^{z_{dim}} \int_{-\infty}^{\infty}  \frac{q(z_k| \boldsymbol{\uptheta}_j^{*}) \ q(z_k| \mathbf{x}_{\text{obs}}^{(i)})}{q(z_k)} \ dz_k
		\end{equation}
		Because the output \(q(\cdot)\) of the encoder follows a Gaussian distribution, it can be expressed as
		\begin{equation}
		\begin{aligned} &L(\mathbf{X}_{\text{obs}}|\boldsymbol{\upphi})  \\
		&=\prod_{i=1}^{N_{obs}} \sum_{j=1}^{N_{sim}} \prod_{k=1}^{z_{dim}} \frac{{}_3^{}\sigma_{(k)}}{\sqrt{2\pi}{}_1^{}\sigma_{(k,i)}{}_2^{}\sigma_{(k,j)}}
		\int_{-\infty}^{\infty}\exp\left( \textstyle -\frac{(z_k-{}_1^{}\mu_{(k,i)})^2}{2{}_1^{}\sigma_{(k,i)}^2}
		-\frac{(z_k-{}_2^{}\mu_{k,j})^2}{2{}_2^{}\sigma_{(k,j)}^2}
		+\frac{(z_k-{}_3^{}\mu_{(k)})^2}{2{}_3^{}\sigma_{(k)}^2} \right) dz_k
		\end{aligned}
		\end{equation}
		Here, \(z_{k{|\mathbf{X}_\text{obs}^{(i)}}}\sim \mathcal{N}({}_1^{}\mu_{(k,i)}^{},{}_1^{}\sigma_{(k,i)}^{})\), \(z_{k{|h(\boldsymbol{\uptheta}_j^{*})}}\sim \mathcal{N}({}_2^{}\mu_{(k,j)}^{},{}_2^{}\sigma_{(k,j)}^{})\), and \(z_{k}\sim \mathcal{N}({}_3^{}\mu_{(k)}^{},{}_3^{}\sigma_{(k)}^{})\), all of which can be obtained from the encoder.
		As the exponential part becomes a quadratic polynomial, it can be further simplified into a Gaussian distribution form, leading to the following result:
		\begin{equation}
			L(\mathbf{X}_{\text{obs}}|\boldsymbol{\upphi}) =
			\prod_{i=1}^{N_{obs}} \sum_{j=1}^{N_{sim}}\prod_{k=1}^{z_{dim}} d_{(i,j,k)} \exp{\left( \frac{b_{(i,j,k)}^2-4a_{(i,j,k)}c_{(i,j,k)}}{4a_{(i,j,k)}}\right)}\sqrt{\frac{\pi}{a_{(i,j,k)}}}
		\end{equation}
		where, $a_{(i,j,k)}=\frac{1}{2{}_1^{}\sigma_{(k,i)}^2}+\frac{1}{2{}_2^{}\sigma_{(k,j)}^2}-\frac{1}{2{}_3^{}\sigma_{(k)}^2}$, $b_{(i,j,k)}=-\frac{{}_1^{}\mu_{(k,i)}^{}}{{}_1^{}\sigma_{(k,i)}^2}-\frac{{}_2^{}\mu_{(k,j)}^{}}{{}_2^{}\sigma_{(k,j)}^2}+\frac{{}_3^{}\mu_{(k)}^{}}{{}_3^{}\sigma_{(k)}^2}$, $c_{k,i,j}=\frac{{}_1^{}\mu_{(k,i)}^2}{2{}_1^{}\sigma_{(k,i)}^2}+\frac{{}_2^{}\mu_{(k,j)}^2}{2{}_2^{}\sigma_{(k,j)}^2}-\frac{{}_3^{}\mu_{(k)}^2}{2{}_3^{}\sigma_{(k)}^2}$, $d_{(i,j,k)}=\frac{{}_3^{}\sigma_{(k)}}{\sqrt{2\pi}{}_1^{}\sigma_{(k,i)}{}_2^{}\sigma_{(k,j)}}$.
		For further details, refer to \cite{Lee23}.
		\clearpage

	\section{Theoretical likelihood of 2DOF example}\label{sec:app1}
		In this Appendix, the theoretical likelihood of the observation matrix $\mathbf{X}_{\text{obs}}$ given hyperparameters $\boldsymbol{\mu}$ and $\sigma$ for the example presented in Section \ref{sec:2dof}, is derived.
		The likelihood of the observation matrix $\mathbf{X}_{\text{obs}}$ given hyperparameters $\boldsymbol{\mu}$ and $\sigma$ can be expressed as the product of the likelihoods of individual observations $\mathbf{x}_{\text{obs}}^{(i)}$:
		\begin{equation}
			p(\mathbf{X}_{\text{obs}}|\boldsymbol{\mu},\sigma) = \prod_{i=1}^{N_{obs}} p(\mathbf{x}_{\text{obs}}^{(i)}|\boldsymbol{\mu},\sigma)
		\end{equation}
		To further expand the likelihood, marginalization over the parameter space $\boldsymbol{\uptheta}$ is performed, considering the conditional probability of the observations, given the parameters and the prior distribution of the parameters:
		\begin{equation}
			p(\mathbf{X}_{\text{obs}}|\boldsymbol{\mu},\sigma) = \prod_{i=1}^{N_{obs}} \int p(\mathbf{x}_{\text{obs}}^{(i)} | \boldsymbol{\uptheta}, \boldsymbol{\mu},\sigma) p(\boldsymbol{\uptheta} | \boldsymbol{\mu},\sigma) \, d\boldsymbol{\uptheta}
		\end{equation}
		Using the delta function, the likelihood can be expressed as:
		\begin{equation}
			p(\mathbf{X}_{\text{obs}}|\boldsymbol{\mu},\sigma) = \prod_{i=1}^{N_{obs}} \int \delta(\mathbf{x}_{\text{obs}}^{(i)} - h(\boldsymbol{\uptheta}))p(\boldsymbol{\uptheta} | \boldsymbol{\mu},\sigma) \, d\boldsymbol{\uptheta}
		\end{equation}
		The delta function $\delta(\mathbf{x}_{\text{obs}}^{(i)} - h(\boldsymbol{\uptheta}))$ enforces the condition that $\mathbf{x}_{\text{obs}}^{(i)}$ equals to $h(\boldsymbol{\uptheta})$.
		This allows us to integrate over the parameter space $\boldsymbol{\uptheta}$, effectively summing over solutions $\boldsymbol{\uptheta}_{i,j}^{\ast}$ that satisfy the condition.
		Using the sifting property of the delta function, the likelihood can be rewritten as
		\begin{equation}
			p(\mathbf{X}_{\text{obs}}|\boldsymbol{\mu},\sigma) = \prod_{i=1}^{N_{obs}} \sum_{j=1}^{N_{sol}} \frac{1}{|h'\left(\boldsymbol{\uptheta}_{i, j}^{\ast} \right)|} p \left( \boldsymbol{\uptheta}_{i, j}^{\ast} | \boldsymbol{\mu},\sigma \right)
		\end{equation}
		where $\boldsymbol{\uptheta}_{i,j}^{\ast}$ is the $j$th true parameter of the $i$th observation.
		$N_{sol}$ represents the number of possible true parameters (i.e., solutions).
		\begin{equation}
			p(\mathbf{x}_{\text{obs}}^{(i)}|\boldsymbol{\mu},\sigma) \propto \prod_{i=1}^{N_{obs}} \sum_{j=1}^{N_{sol}} \frac{1}{|h'(\boldsymbol{\uptheta}_{i,j}^{\ast})|} \exp\left(-\frac{\|\boldsymbol{\uptheta}_{i,j}^{\ast}-\boldsymbol{\mu}\|_2^2}{2\sigma^2}\right)
		\end{equation}
		In this case, because $h$ is a one-to-one correspondence, assuming $N_{sol}=1$, we obtain
		\begin{equation}
			p(\mathbf{x}_{\text{obs}}^{(i)}|\boldsymbol{\mu},\sigma) \propto \prod_{i=1}^{N_{obs}} \frac{1}{|h'(\boldsymbol{\uptheta}_{i}^{\ast})|} \exp\left(-\frac{\|\boldsymbol{\uptheta}_{i}^{\ast}-\boldsymbol{\mu}\|_2^2}{2\sigma^2}\right)
		\end{equation}
		Simplifying this further, we obtain
		\begin{equation}
			p(\mathbf{x}_{\text{obs}}^{(i)}|\boldsymbol{\mu},\sigma) \propto \prod_{i=1}^{N_{obs}} \exp\left(-\frac{\|\boldsymbol{\uptheta}_{i}^{\ast}-\boldsymbol{\mu}\|_2^2}{2\sigma^2}\right)
		\end{equation}

	\section{Samples of sequences from the  maximum a posteriori estimation at NASA challenge}\label{sec:app2}
		\begin{figure}[H]
			\begin{tabular}{cc}
				\begin{minipage}[t]{0.48\hsize}
					\centering
					\includegraphics[width=0.98\linewidth]{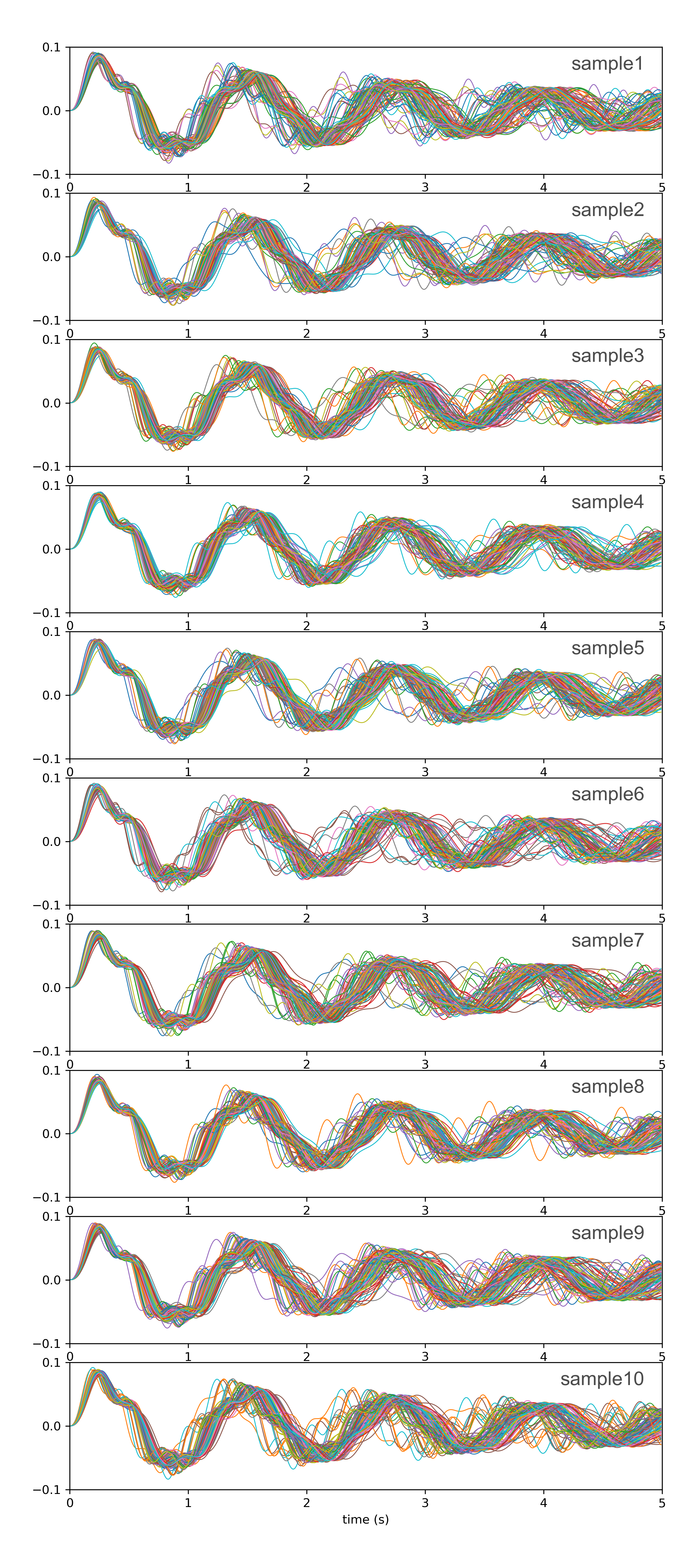}
				\end{minipage}
				\begin{minipage}[t]{0.48\hsize}
					\centering
					\includegraphics[width=0.98\linewidth]{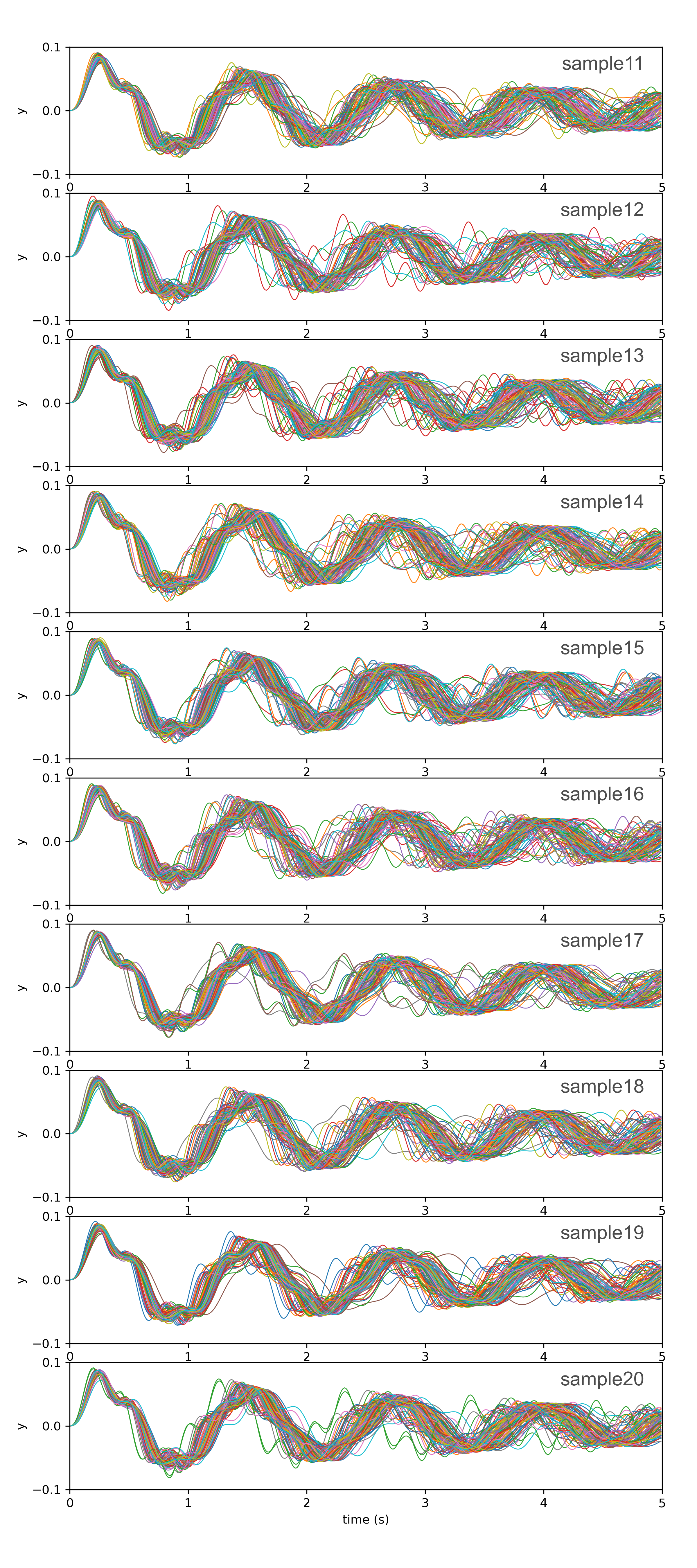}
				\end{minipage}\\
				\\
			\end{tabular}
			\caption{Twenty sample sequences obtained using the maximum likelihood hyperparameters}\label{}
		\end{figure}
		\clearpage

		\bibliographystyle{elsarticle-num}
		\bibliography{main}

\end{document}